\documentclass[12pt]{article}
\usepackage[utf8]{inputenc}
\usepackage{amsmath, amssymb, amsthm, bbm, mathrsfs}
\usepackage{hyperref}
\hypersetup{
colorlinks,
citecolor=blue,
filecolor=blue,
linkcolor=blue,
urlcolor=blue
}
\usepackage{geometry}
\usepackage{natbib}
\usepackage{titling}
\usepackage{authblk}
\usepackage[linesnumbered,ruled,vlined]{algorithm2e}

\SetKwInput{KwInput}{Input}               
\SetKwInput{KwOutput}{Output}  
\SetKwInput{KwSet}{Set}             
\SetKwProg{Fn}{Function}{:}{}

\usepackage{verbatim}
\usepackage{multirow}
\usepackage{booktabs, array, threeparttable}
\usepackage{rotating}
\usepackage{graphicx}
\usepackage{subfigure}
\usepackage{float}
\usepackage{caption}
\usepackage{amsfonts}
\usepackage{xr}
\usepackage{color}
\usepackage{enumerate}
\usepackage{epstopdf}
\usepackage{ulem}
\usepackage{makecell}
\allowdisplaybreaks
\graphicspath{{./figs/}}

\makeatletter
\newcommand*{\addFileDependency}[1]{
  \typeout{(#1)}
  \@addtofilelist{#1}
  \IfFileExists{#1}{}{\typeout{No file #1.}}
}
\makeatother

\newcommand*{\myexternaldocument}[1]{
    \externaldocument{#1}
    \addFileDependency{#1.tex}
    \addFileDependency{#1.aux}
}

\myexternaldocument{PFL_TIR_supp}

\usepackage{geometry}
\newcommand{\T}{{\top}}
\newcommand{\cA}{{\mathcal{A}}}

\newcommand{\cL}{{\mathcal{L}}}

\newcommand{\cS}{{\mathcal{S}}}

\newcommand{\cD}{{\mathcal{D}}}

\newcommand{\bbE}{{\mathbb{E}}}
\newcommand{\bbP}{\mathbb{P}}

\newcommand{\commenting}[1]{}

\newcommand{\bbeta}{{\boldsymbol{\beta}}}

\newcommand{\bth}{\boldsymbol{\theta}}
\newcommand{\bde}{\boldsymbol{\delta}}

\newcommand{\cG}{{\mathcal{G}}}

\newcommand{\bx}{{\boldsymbol{x}}}

\newtheorem{theorem}{Theorem}[section]

\newtheorem{remark}{Remark}[section]
\newtheorem{corollary}{Corollary}[section]
\newtheorem{assumption}{Assumption}[section]

\newtheorem{proposition}{Proposition}[section]

\title{Efficient Federated Estimation and Inference for High-Dimensional Tail Index Regression}
 \date{}
\author{Haoyu Geng$^1$,\ Liuhua Peng$^2$,\ Changliang Zou$^1$ and Xiaolong Cui$^{3}$\footnote{Corresponding author: nk.xlcui@gmail.com. All authors equally contributed to this work.}
\ \                   \\
{$^1${\small\it School of Statistics and Data Science, Nankai University, China}}\\
{$^2${\small\it School of Mathematics and Statistics, The University of Melbourne, Australia}}\\
{$^3${\small\it School of Statistics and Data Science, Shanghai University of Finance and Economics, China}}
}

\begin{document}
\setlength{\droptitle}{-2cm}
\maketitle

\vspace{-2cm}
\begin{abstract}
\baselineskip 20pt

Tail index regression studies how covariates affect tail heaviness in heavy-tailed data. In many applications, data are distributed across heterogeneous sources, where direct pooling is infeasible due to privacy or regulatory constraints. Existing methods mainly focus on single-dataset analysis and do not address heterogeneous federated settings. We develop a personalized federated framework for high-dimensional tail index regression that accommodates client heterogeneity while exploiting latent similarities across clients. The proposed estimator combines sparsity regularization with nonconcave fusion penalties to perform coefficient estimation, variable selection, and group recovery. We establish non-asymptotic convergence rates and show that the estimator enjoys an oracle property by consistently recovering the underlying grouping structure. For computation, we develop an ADMM-based federated algorithm with adaptive gradient updates and establish its convergence guarantees. We further propose a debiased federated inference procedure based on adaptive weighted aggregation across related clients, yielding valid confidence intervals and hypothesis tests with improved efficiency over target-only inference. Simulation studies and real-data analysis demonstrate the effectiveness of the proposed methods.

\end{abstract}
\noindent{\bf Keywords}: Heterogeneous data; fusion penalty; sparse estimation; ADMM algorithm; debiased inference.

\renewcommand{\baselinestretch}{1.5}

\section{Introduction}

Heavy-tailed data arise frequently in finance, insurance, environmental studies, and other domains where rare but severe events are of primary concern \citep{resnick2007heavy}. A central quantity in the analysis is the tail index, or extreme value index \citep{de2006extreme}, which governs the decay rate of tail probabilities and hence the frequency and severity of extreme outcomes. Classical estimators, such as the Hill and Pickands estimators \citep{pickands1975statistical,hill1975simple}, focus on the marginal tail behavior of a response variable. In modern applications, observations are often accompanied by a large amount of covariates, such as policyholder characteristics in insurance or firm- and market-level attributes in finance, that may systematically affect tail behavior. This shifts the focus from estimating a marginal tail index to understanding how covariates affect conditional tail heaviness.

Tail index regression \citep{beirlant2003regression,wang2009tail} addresses this problem by modeling the tail index of a response $y$ as a function of covariates $\bx\in\mathbb{R}^p$. A common specification assumes a Pareto-type conditional tail model with $\alpha(\bx)=\exp(\bx^\top\bth^\ast)$, where $\bth^\ast=(\theta^\ast_1,\ldots,\theta^\ast_p)^\top$ is the regression coefficient. Under this model, $\theta^\ast_j$ quantifies the effect of the $j$-th covariate on the tail index, and hence on the heaviness of the conditional tail. Estimation and inference for these coefficients are therefore essential for identifying covariates that drive extreme risks and quantifying the uncertainty of their effects.

Despite its importance, inference for tail index regression remains challenging. Tail-based estimation and inference use only observations exceeding a high threshold, so the effective sample size can be much smaller than the nominal sample size, a fundamental bottleneck in extreme value theory \citep{de2006extreme}. The problem is further amplified in high-dimensional settings, where the number of covariates may be comparable to or larger than the number of tail observations. Although recent work has developed regularized estimation and debiased inference for high-dimensional tail index regression \citep{sasaki2024high}, inference based on a single dataset often suffers from low statistical efficiency, producing highly variable estimates and wide confidence intervals with limited inferential value.

In many applications, the target dataset is accompanied by related datasets from other sources, such as loss records from different insurers or extreme-event data from different financial institutions. Although these data cannot usually be directly pooled because of privacy, ownership, or regulatory constraints, they may contain useful tail information. This motivates federated learning, which enables information sharing without transferring raw observations \citep{kairouz2021advances}. However, federated tail index regression cannot rely on naive aggregation. Auxiliary clients may differ from the target client in their covariate distributions and in their covariate effects on tail behavior. Borrowing information from incompatible clients may therefore introduce bias, while ignoring related clients loses efficiency. The key challenge is to borrow strength selectively from informative clients while protecting target-specific inference from heterogeneity-induced bias.

In this paper, we develop a \textit{personalized federated framework} \citep{tan2022towards,liu2025robust} for high-dimensional tail index regression, with the goal of estimating and making inference on covariate effects on the tail index under heterogeneous federated data. For each client $k\in[K]$, we assume a Pareto-type conditional tail model with tail index $\alpha_k(\bx)=\exp(\bx^\top\bth_k^\ast)$, where $\bth_k^\ast=(\theta^\ast_{k1},\ldots,\theta^\ast_{kp})^\top$ is sparse and client-specific. To capture latent similarities across clients, we allow each coefficient $\theta^\ast_{kj}$ to vary across clients but assume that, for each covariate $j$, the corresponding coefficients may form unknown groups. This coefficient-level grouping structure enables personalized information sharing: clients with similar effects for a given covariate can be fused, while heterogeneous effects remain separated. Such targeted collaboration is particularly important for extreme value analysis, where the limited number of tail observations makes auxiliary information valuable.

The contributions of this paper are as follows.     
First, we propose a personalized federated tail index regression framework for heterogeneous heavy-tailed data. Unlike existing distributed tail inference methods that ignore covariates or impose a common tail structure, our framework allows client-specific coefficients while exploiting latent similarity through clustered heterogeneity. Combining sparsity regularization with nonconcave fusion penalties, the method performs variable selection, coefficient estimation, and group recovery simultaneously. We establish non-asymptotic convergence rates and show that the estimator enjoys an oracle property, asymptotically recovering the true grouping structure and achieving efficiency gains over target-only estimators. 
A key theoretical novelty is that the analysis must handle the Pareto-type tail approximation error inherited from extreme value theory, which makes the problem fundamentally different from standard high-dimensional M-estimation. In particular, we employ advanced concentration inequalities and tools from empirical process theory, including the symmetrization theorem, contraction theorem and peeling argument, to characterize the second-order curvature of the likelihood function, which is crucial for establishing non-asymptotic convergence and oracle properties. Such theoretical techniques may be of independent interest for high-dimensional extreme value inference.

Second, we develop an Alternating Direction Method of Multipliers (ADMM)-based algorithm implemented in a federated client-server framework for computing the proposed estimator. To handle nonconcave penalties and loss functions with non-Lipschitz gradients, for which existing convergence theory does not apply, we integrate adaptive gradient updates \citep{berahas2024non} within the ADMM iterations. The resulting method admits efficient federated implementation that preserves privacy and communication efficiency. By exploiting the local curvature of the loss function, we establish theoretical guarantees for the convergence of the proposed algorithm.

Third, we develop a federated inference procedure for regression parameters under the personalized framework. 
The proposed approach combines debiasing with adaptive aggregation across related datasets, leading to valid confidence intervals and hypothesis tests. We show that the proposed inferential procedure is more efficient than inference based on the target dataset alone. 
In particular, it achieves smaller asymptotic variance and produces tighter confidence intervals, demonstrating that personalized information sharing improves both estimation and inference efficiency in high-dimensional tail index regression.

\subsection{Related works}

Tail index regression was introduced by \citet{beirlant2003regression,wang2009tail}, who proposed a parametric regression model for the tail index using a log-link specification.
Subsequent works mainly focused on fixed-dimensional settings, including semiparametric models \citep{li2022semiparametric}, varying-coefficient models \citep{momoki2024hypothesis}, single-index models \citep{yoshida2025single}, and time-series tail index regression \citep{nicolau2023tail}. Recently, \citet{sasaki2024high} studied regularized estimation and debiased inference for high-dimensional tail index regression. However, their framework considers a single dataset and does not address federated learning or heterogeneous information sharing across clients.

Under the distributed setting, \citet{chen2022distributed} studied inference for a common extreme value index without covariates, showing that decentralized aggregation can improve efficiency. However, the common-tail-index assumption can be restrictive when auxiliary datasets differ from the target. More recently, \citet{chen2024high} developed high-dimensional procedures for testing equality of extreme value indices in multivariate extremes. These works advance distributed tail inference, but they do not address covariate-dependent tail index regression, target-specific coefficient heterogeneity, or selective information sharing across clients. In contrast, we develop a personalized federated framework for high-dimensional tail index regression that accommodates heterogeneous client structures and provides both federated estimation and valid debiased inference for covariate effects.

Personalized federated learning constructs client-specific models and is particularly suitable when local data distributions are heterogeneous. Existing approaches include fine-tuning \citep{fallah2020personalized}, multi-task learning \citep{smith2017federated,huang2021personalized}, clustered federated learning \citep{ghosh2020efficient,sattler2021clustered}, parameter decoupling \citep{arivazhagan2019federated}, and knowledge distillation \citep{li2019fedmd}. Recently, fusion-penalty methods have been developed to capture structured heterogeneity in high-dimensional federated models. For example, \citet{liu2025robust} studied robust personalized federated regression with sparse fused regularization; \citet{li2025personalized} incorporated sparse orthogonal factor learning and fusion penalization for large-scale association networks; and \citet{zhou2026row} proposed a sparse row-wise fusion regularizer for multivariate personalized federated learning. However, personalized federated methods for high-dimensional tail index regression remain unexplored.

The rest of the paper is organised as follows. 
Section~\ref{sec:FD_problem} introduces the personalized federated tail index regression framework, including the model formulation and penalized estimation procedure, Section~\ref{sec:theory} presents the theoretical properties, and the proposed federated ADMM algorithm with adaptive gradient updates. Section~\ref{sec:FD_inference} develops personalized federated inference procedures based on weighted debiasing and establishes the asymptotic properties of the proposed estimators. Section~\ref{sec:simu} evaluates the finite-sample performance of the proposed methods through extensive simulation studies. Section~\ref{sec:real_data} presents a real-data application. Technical proofs, auxiliary lemmas, and additional numerical results are collected in the supplementary material.

\textbf{Notations.}
For an integer $n$, let $[n]=\{1,2,\dots,n\}$. 
For a set $\cA$, $|\cA|$ denotes its cardinality. 
For any two real numbers $a$ and $b$, we denote $a \vee b = \max(a,b)$ and $a \wedge b = \min(a,b)$. 
For a vector $\bx=(x_1,\ldots,x_p)^{\top}\in\mathbb{R}^p$, we define its $\ell_0$, $\ell_1$, $\ell_2$, and $\ell_{\infty}$ norms as 
$\|\bx\|_0=\sum_{j=1}^p \mathbbm{1}\{x_j\neq 0\}$, 
$\|\bx\|_1=\sum_{j=1}^p |x_j|$, 
$\|\bx\|_2=\sqrt{\sum_{j=1}^p x_j^2}$, and 
$\|\bx\|_{\infty}=\max_{1\le j\le p}|x_j|$, respectively. 
Let $\mathbb{S}^{p-1} = \{\bx \in \mathbb{R}^p : \|\bx\|_2 = 1\}$ denote the unit sphere in $\mathbb{R}^p$. 
Let $\boldsymbol{e}_{j,p}$ denote the $j$-th canonical basis vector in $\mathbb{R}^p$. 
For a matrix $\mathbf{A}=(a_{ij})$, its elementwise $\ell_{\infty}$ norm is defined as 
$\|\mathbf{A}\|_{\infty}=\max_{i,j}|a_{ij}|$. 
If $\mathbf{A}$ is a square matrix, we denote its largest and smallest eigenvalues by 
$\lambda_{\max}(\mathbf{A})$ and $\lambda_{\min}(\mathbf{A})$, respectively. 
Let $\mathbf{I}_{p \times p}$ denote the $p\times p$ identity matrix.

\section{Model and personalized federated estimation}\label{sec:FD_problem}

We consider a personalized federated learning framework with $K$ clients. For each client $k\in[K]$, let $\cD_k=\{(\bx_i^{(k)},y_i^{(k)})\}_{i=1}^{N_k}$ denote the local dataset, where $\bx_i^{(k)}=\big(x_{i1}^{(k)}, x_{i2}^{(k)}, \ldots, x_{ip}^{(k)}\big)^{\top}$ is the $p$-dimensional covariate vector and $y_i^{(k)}$ is a heavy-tailed response. 
We assume that all data points are independent both within each client and across clients, while allowing the covariate distributions $\bx^{(k)}$ to differ across clients.
For client $k$, we assume that the conditional tail of $y^{(k)}$ given $\bx^{(k)}$ follows the Pareto-type model
\begin{align}\label{Pareto-type tail model}
    \mathbb{P}(y^{(k)} > t \mid \bx^{(k)}) = t^{-\alpha_k(\bx^{(k)})} \mathcal{L}_k(t; \bx^{(k)}),
\end{align}
where $\alpha_k(\bx^{(k)}) = \exp\left(\bx^{(k)\top} \bth_k^*\right)$ is the client-specific conditional tail index, which is assumed to be bounded throughout the article, i.e., $\overline{\alpha}\geq \alpha_k(\bx^{(k)})\geq \underline{\alpha} > 0$. In addition, $\mathcal{L}(t; \bx)$ is a covariate-dependent slowly varying function satisfying $\mathcal{L}_k(th; \bx^{(k)})/\mathcal{L}_k(t; \bx^{(k)}) \to 1,$ for any $h > 0$ as $t\to\infty$.
The parameter $\boldsymbol{\theta}_k^* = (\theta_{k1}^*, \theta_{k2}^*,\ldots, \theta_{kp}^*)^{\top}$ represents the $p$-dimensional true \textit{sparse} regression coefficient vector so that only a small subset of the covariates $\bx^{(k)}$ affects the tail behavior of $y^{(k)}$.
Moreover, $\bth_k^\ast$'s are allowed to differ across clients, reflecting heterogeneity in tail behaviour.

Adapting the approach of \citet{hall1982some}, we consider the Hall class of heavy-tailed distributions, under which $\mathcal{L}_k(t;\bx^{(k)})$ satisfies
\begin{align}\label{Hall class}
   \mathcal{L}_k(t;\bx^{(k)})
= c_{0k}(\bx^{(k)}) + c_{1k}(\bx^{(k)})t^{-\beta_k(\bx^{(k)})} + r_k(t,\bx^{(k)})
\quad \text{as } t \to \infty, 
\end{align}
for some functions $\beta_k(\bx^{(k)}) \geq \underline{\beta} > 0$, $c_{0k}(\bx^{(k)}) \geq \underline{c_0}> 0$, $|c_{1k}(\bx^{(k)})|\leq \overline{c_1} $, and a remainder term $r_k(t,\bx^{(k)})$ such that
$\sup_k\sup_{\bx^{(k)}} |r_k(t,\bx^{(k)})t^{\beta_k(\bx^{(k)})}| \to 0$ as $t \to \infty$.

Under model~\eqref{Pareto-type tail model}, estimation of $\bth_{k}^*$ and the tail index $\alpha_k(\bx^{(k)})$ is based on exceedances over a high threshold, since the Pareto-type approximation is valid only in the tail region. For client $k$, let $w_k$ be a high threshold and define the effective tail sample size $n_k=\sum_{i=1}^{N_k}\mathbbm{1}\{y_i^{(k)}>w_k\}$.
Conditioning on $\{y_i^{(k)}>w_k\}$ and $\bx_i^{(k)}$, $\{y_i^{(k)}: y_i^{(k)}>w_k\}$ are independent and the conditional distribution
$f_{y_i^{(k)}\mid y_i^{(k)}>w_k,\bx_i^{(k)}}(t)\approx \alpha_k(\bx_i^{(k)})(t/w_k)^{-\alpha_k(\bx_i^{(k)})}t^{-1}$ for large $w_k$.
Without loss of generality, we can order the data such that $y_i^{(k)}>w_k$ for $i=1,\dots,n_k$.

Locally on each client, $\bth_k^*$ can be estimated by minimizing the following approximate negative log-likelihood function \citep{wang2009tail,nicolau2023tail}: 
\begin{equation}\label{individual_problem}
    \widehat{\bth}_k^{\text{local}} = \underset{\boldsymbol{\theta}_k \in \mathbb{R}^{p}}{\arg \min }\ \frac{1}{n_k} \sum_{i=1}^{n_k}  \ell\big(y_{i}^{(k)},\bx_{i}^{(k)},\boldsymbol{\theta}_k\big),
\end{equation}
where 
$\ell\big(y_{i}^{(k)},\bx_{i}^{(k)},\boldsymbol{\theta}_k\big) = \exp{(\bx_{i}^{(k)\T}\bth_k)}\log(y_{i}^{(k)}/w_{k}) - \bx_{i}^{(k)\T}\bth_k$.
In high-dimensional settings, \citet{sasaki2024high} developed regularized estimation by adding an $\ell_1$ penalty on~\eqref{individual_problem}.

A key challenge is that only observations exceeding the threshold contribute to estimation. Consequently, even when the original sample size $N_k$ is large, the effective sample size $n_k$ may remain relatively small. This issue becomes substantially more severe in high-dimensional settings, where reliable estimation and inference require sufficient tail observations relative to the number of covariates. As a result, local estimation based on a single client may suffer from high variability and low inferential efficiency, often producing unstable estimates and wide confidence intervals.

In practice, the coefficient vectors $\bth_k^*$ often share structural similarities across clients. Some client-specific coefficients may share common values for certain covariates, such as $\theta_{kj}^* = \theta_{k'j}^*$, which means that the $j$-th covariate has the same effect for clients $k$ and $k'$. Such latent similarities provide an opportunity for personalized information sharing: each client retains its own regression model while borrowing strength from similar clients in selected components. This motivates the following coordinate-wise grouping structure.

\begin{assumption}\label{assumption: parameter group structure}
    For each $j\in[p]$, the $K$ coefficients $(\theta_{1 j}^*, \theta_{2 j}^*, \ldots, \theta_{K j}^*)$  are grouped into $M_j$ groups $\mathcal{G}_1^j, \mathcal{G}_2^j, \ldots \mathcal{G}_{M_j}^j$ such that $\mathcal{G}_i^j \cap \mathcal{G}_{i^{\prime}}^j=\varnothing$ for $i \neq i^{\prime}$ and $\bigcup_{i=1}^{M_j} \mathcal{G}_i^j=[K]$.
    Moreover, coefficients are identical within each group: for any $i\in[M_j]$ and any $k,k'\in\mathcal{G}_i^j$,
    $\theta_{kj}^*=\theta_{k'j}^*$.
\end{assumption}

To leverage these shared patterns while accommodating client-specific heterogeneity, we adopt a personalized federated learning framework \citep{liu2025robust}. For notational convenience, we vectorize
the $K$ coefficients as $\bth^*=(\bth_1^{*\T},\bth_2^{*\T},\dots,\bth_K^{*\T})^{\T}\in \mathbb{R}^{pK}$. The proposed the personalized federated estimator (PFE) $\widehat{\boldsymbol{\theta}}=(\widehat{\boldsymbol{\theta}}_1^{\T},\widehat{\boldsymbol{\theta}}_2^{\T},\dots,\widehat{\boldsymbol{\theta}}_K^{\T})^{\T}\in \mathbb{R}^{pK}$ is
\begin{equation}\label{FD_estimator}
    \widehat{\boldsymbol{\theta}}= \underset{\boldsymbol{\theta} \in \mathbb{R}^{pK}}{\arg \min }\ \frac{1}{n} \sum_{k=1}^K \sum_{i=1}^{n_k} \ell\big(y_{i}^{(k)},\bx_{i}^{(k)},\boldsymbol{\theta}_k\big)  
    + \frac{1}{K}\sum_{j=1}^{p}\sum_{k=1}^Kp_{\lambda_1}(|\theta_{kj}|)
+ \frac{1}{K}\sum_{j=1}^{p}\sum_{k\leq k'}p_{\lambda_2}(|\theta_{kj}-\theta_{k'j}|),
\end{equation}
where $\boldsymbol{\theta}=(\boldsymbol{\theta}_1^{\T},\ldots,\boldsymbol{\theta}_K^{\T})^{\T}$, $n=\sum_{k=1}^K n_k$ denotes the total effective sample size, and $p_\lambda(\cdot)$ is a penalty function with a tuning parameter $\lambda \geq 0$. 
The penalization in~\eqref{FD_estimator} addresses both high dimensionality and cross-client heterogeneity. The first term induces sparsity in each client-specific coefficient vector, while the second fuses similar covariate effects across clients. The tuning parameter $\lambda_2$ controls the degree of personalized information sharing.

We focus on penalty functions satisfying the following conditions.

\begin{assumption}\label{assumption: penalty conditions}
(i) The function $p_\lambda(z)$ satisfies $p_\lambda(0)=0$ and is symmetric around zero;
(ii) For $z>0$, the function $p_\lambda(z)$ is nondecreasing, and $p_\lambda(z)/z$ is nonincreasing in $z$;
(iii) The function $p_\lambda(z)$ is differentiable for all $z \neq 0$ and subdifferentiable at $z=0$, with $\lim _{z \rightarrow 0^{+}} p_\lambda^{\prime}(z)=\lambda$;
(iv) There exists $v>0$ such that $p_{\lambda, v}(z):=p_\lambda(z)+ vz^2/2$ is convex;
(v) There is some scalar $a \in(0, \infty)$ such that $p_\lambda^{\prime}(z)=0$ and $p_\lambda(z)\geq g(a,\lambda)>0$, for all $z \geq a \lambda$ and some nondecreasing function $g(a,\lambda)$.
\end{assumption}

Assumption~\ref{assumption: penalty conditions} is satisfied by many commonly used penalty functions~\citep{loh2015regularized}. Two important nonconcave examples are SCAD~\citep{fan2001variable} and MCP~\citep{zhang2010nearly}, both of which are shown to outperform $\ell_1$ penalty in recovering sparse structures and reducing estimation bias.
The SCAD and MCP penalties are defined as
\[
p^{\text{SCAD}}_\lambda(z) =
\begin{cases}
\lambda |z| & \text{if } |z| \leq \lambda, \\
\frac{ 2a\lambda|z|-\lambda^2-z^2}{2(a-1)} & \text{if } \lambda < |z| \leq a\lambda, \\
\frac{(a+1)\lambda^2}{2} & \text{if } |z| > a\lambda,
\end{cases}
\qquad
p^{\text{MCP}}_\lambda(z) =
\begin{cases}
\lambda |z| - \frac{z^2}{2a} & \text{if } |z| \leq a \lambda, \\
\frac{a\lambda^2}{2} & \text{if } |z| > a \lambda.
\end{cases}
\]
For SCAD, $a>2$ and the conditions in Assumption \ref{assumption: penalty conditions} hold with $v=1/(a-1)$ and $g(a,\lambda)=(a+1)\lambda^2/2$. For MCP, $a>1$ and the conditions in Assumption \ref{assumption: penalty conditions} hold with $v=1/a$ and $g(a,\lambda)=a\lambda^2/2$.

Compared with \citet{sasaki2024high}, which studies high-dimensional tail-index regression using an $\ell_1$ penalty under an exact Pareto tail model, we consider a more general and challenging personalized federated framework with latent coordinate-wise grouping structures under the more general Pareto-type tail model~\eqref{Pareto-type tail model}. The additional slowly varying component introduces substantially more challenging theoretical analysis.

\section{Theoretical properties and federated algorithm}\label{sec:theory}

\subsection{Theoretical properties}

In this section, we study the theoretical properties of the proposed PFE estimator in~\eqref{FD_estimator}. Since the true grouping structure plays a central role in the analysis, we first introduce an oracle reparametrization that represents the distinct coefficient values after grouping.

For each $j\in[p]$, let $\beta_i^{j*}$ denote the common value in group $\cG_i^j$, namely, $\theta_{kj}^*=\beta_i^{j*} $ for all $k\in\cG_i^j$ and $i\in[M_j]$.
Denote
$\bbeta^{j*}=(\beta_1^{j*},\ldots,\beta_{M_j}^{j*})^{\T}$
as the vector of distinct coefficient values for the $j$-th covariate, and stack these vectors as
$ \bbeta^*=(\bbeta^{1*\T},\ldots,\bbeta^{p*\T})^{\T}\in\mathbb{R}^M, $
where $M=\sum_{j=1}^p M_j$.
Thus, $\bbeta^*$ collects all distinct coefficients implied by the latent grouping structure. For convenience, we also write
$\bbeta^*\triangleq(\beta_1^*,\ldots,\beta_M^*)^{\T}\in\mathbb{R}^M$.

To link the oracle parameter $\bbeta^*$ with the original parameter
$\bth^*=(\bth_1^{*\T},\ldots,\bth_K^{*\T})^{\T}\in\mathbb{R}^{pK}$,
define the membership matrix $\mathbf U\in\mathbb{R}^{pK\times M}$. Let $\mathbf U_i=(U_{i1},\ldots,U_{iM})$ be the $i$-th row of $\mathbf U$. If the $i$-th component of $\bth^*$ corresponds to the $j$-th component of $\bbeta^*$, then $U_{ij}=1$ and $U_{ik}=0$ for all $k\neq j$. Thus, each row of $\mathbf U$ contains a single nonzero entry, and $\bth^*=\mathbf U\bbeta^*$. Let $\mathbf U_k\in\mathbb{R}^{p\times M}$ be the submatrix of $\mathbf U$ corresponding to client $k$, so that $\bth_k^*=\mathbf U_k\bbeta^*$.

For notational convenience, we relabel the groups
$\{\cG_i^j: i\in[M_j], j\in[p]\}$ as $\{\cG_\ell\}_{\ell=1}^M$, with $\cG_\ell=\cG_i^j$ when $\ell=\sum_{t=1}^{j-1}M_t+i$.
Let $\cG_{\max}=\max_{\ell\in[M]}|\cG_\ell|$ and $\cG_{\min}=\min_{\ell\in[M]}|\cG_\ell|$ denote the largest and smallest group sizes, respectively.
We also define $n_i^j=\sum_{k\in\cG_i^j}n_k$ as the effective sample size for group $\cG_i^j$, $n_{\min}^{\operatorname{group}}=\min_{i,j} n_i^j$ as the minimal group-wise effective size, and $n_{\min}^{\operatorname{client}}=\min_k n_k$, $n_{\max}^{\operatorname{client}}=\max_k n_k$ as the minimal and maximal client-wise effective sizes.

If the group structures $\{\cG_\ell\}_{\ell=1}^M$ are known, the oracle estimator is defined as
\begin{equation}\label{oracle estimator}
\widehat{\bbeta}
= \underset{\bbeta \in \mathbb{R}^{M}}{\arg \min }\ \frac{1}{n} \sum_{k=1}^K  \sum_{i=1}^{n_k} \ell\big(y_{i}^{(k)},\bx_{i}^{(k)},\mathbf{U}_k\bbeta\big)  
+ \frac{1}{K}\sum_{\ell=1}^{M}|\mathcal{G}_\ell|p_{\lambda_1}(|\beta_{\ell}|).
\end{equation}
This oracle estimator serves as a benchmark for the proposed PFE estimator $\widehat{\boldsymbol{\theta}}$ in~\eqref{FD_estimator}.

Let $\cS=\{y_i^{(k)}>w_k: i\in[N_k],\, k\in[K]\}$ denote the tail sample selection event. The full-sample analysis of \citet{wang2009tail} involves additional complications due to the random ratios $n_k/N_k$. Following \citet{nicolau2023tail}, we instead conduct the theoretical analysis conditional on $\cS$, which avoids these complications while focusing directly on the effective tail samples.
We next introduce assumptions on the model parameters and covariates.

\begin{assumption}\label{assumption: parameter and covariate}
    (i) The parameter $\bbeta^*$ is sparse with $s_*=\|\bbeta^*\|_0$ nonzero elements and satisfies $\|\bbeta^*\|_1\leq C_{\bbeta^*}$ for some constant $C_{\bbeta^*}>0$;
    (ii) For all $k\in[K]$ and $i\in[n_k]$, the covariates $\bx_i^{(k)}$ are uniformly bounded with $\|\bx_i^{(k)}\|_{\infty} \leq C_{\bx}$, and satisfy $\sup_{\bde \in \mathbb{S}^{p-1}} \bbE \big\{|\bde^{\T}\bx_i^{(k)}|^4\mid y_i^{(k)}>w_k\big\} \le C_{\bx}$ for some constant $C_{\bx} > 0$;
    (iii) Let $\boldsymbol{\Sigma}^{(k)}_{w_k}=\bbE\big\{\bx_i^{(k)}\bx_i^{(k)\T}\mid y_i^{(k)}>w_k\big\}$. Assume that its eigenvalues satisfy $C_{\boldsymbol{\Sigma}}^{-1}\leq \lambda_{\min}(\boldsymbol{\Sigma}^{(k)}_{w_k})\leq\lambda_{\max}(\boldsymbol{\Sigma}^{(k)}_{w_k})\leq C_{\boldsymbol{\Sigma}}$ for some constant $C_{\boldsymbol{\Sigma}}>0$ and all $k\in[k]$;
    (iv)  The minimal group-wise effective sample size satisfies $n_{\min}^{\operatorname{group}}/n\geq C_{n_{\min}}$ for some constant $C_{n_{\min}}>0$;
    (v) Let $w_{\min}=\min_{k\in[K]} w_k$. Conditional on $\cS$, assume that $w_{\min}\rightarrow\infty$ as $n=\sum_{k=1}^K\sum_{i=1}^{N_k}\mathbbm{1}(y_i^{(k)}>w_k)\rightarrow \infty$, and $w_{\min}^{-\underline{\beta}}\leq C_w\sqrt{(\log M)/n}$ for some constant $C_w>0$.
\end{assumption}

Assumption~\ref{assumption: parameter and covariate} is standard in the high-dimensional regression and extreme value literature. Assumption \ref{assumption: parameter and covariate}(i) imposes sparsity on the oracle parameter. Assumptions \ref{assumption: parameter and covariate}(ii) and \ref{assumption: parameter and covariate}(iii) impose regularity conditions on the covariates to ensure bounded moments and a well-conditioned covariance matrix under tail sampling \citep{sasaki2024high}, while allowing heterogeneous covariate distributions across clients. Assumption \ref{assumption: parameter and covariate}(iv) prevents any group from having too few effective tail observations. Assumption \ref{assumption: parameter and covariate}(v) is a standard threshold condition under the Hall class, ensuring that the bias of the tail approximation is asymptotically negligible relative to the estimation error.

Let $d_{\min}=\min_{j:|M_j|\geq 2}\min_{ i\neq i'}|\beta^{j*}_i-\beta^{j*}_{i'}|$ denotes the smallest difference between any two distinct group-specific coefficients across all covariates. Denote
$$\varrho_n=\sqrt{\frac{n_{\max}^{\operatorname{client}}}{n}}\sqrt{\frac{\log (pK)}{n}}\vee \sqrt{\frac{\log M}{n}}\vee \frac{n_{\max}^{\operatorname{client}}}{n}\frac{\cG_{\max}^2}{\cG_{\min}^2}\sqrt{\frac{s_*^2\log M}{n}}.$$
The following theorem establishes the oracle property of the proposed PFE estimator. 

\begin{theorem}[Oracle property of estimator $\widehat{\bth}$]\label{thm: oracle property}
    Suppose that Assumption \ref{assumption: parameter group structure}, Assumption \ref{assumption: penalty conditions}, and Assumption \ref{assumption: parameter and covariate} hold.  Let $\lambda_2=\lambda_1+c(K/\cG_{\min})\varrho_n$ and $\lambda_1=c(K/\cG_{\min})\sqrt{(\log M)/n}$ for some conatant $c>0$. Assume that $d_{\min}>a\lambda_2+4C(\cG_{\max}/\cG_{\min})\sqrt{(s_*\log M)/n}$ and $n\gg (\cG_{\max}^4/\cG_{\min}^4)s_*^2\log M$ for some conatant $C>0$. Then conditional on event $\cS$, there exists a local minimizer $\widehat{\bth}$ of the proposed objective function \eqref{FD_estimator} such that $\widehat{\bth}=\mathbf{U}\widehat{\bbeta}$ with probability approaching one as $n\rightarrow\infty$.
\end{theorem}

Theorem~\ref{thm: oracle property} shows that the proposed PFE estimator $\widehat{\bth}$ coincides with the oracle estimator $\widehat{\bbeta}$ with high probability. The condition on $d_{\min}$ ensures that the minimum separation between distinct group-specific coefficients dominates the estimation error. The additional term $\varrho_n$ in the choice of $\lambda_2$ captures the error induced by the unknown group structure and client heterogeneity. The following theorem establishes the convergence rate and group structure consistency of the proposed PFE estimator.

\begin{theorem}[Convergence rate and group structure consistency of $\widehat{\bth}$]\label{thm: rate of theta}
    Suppose that the assumptions of Theorem \ref{thm: oracle property} hold.\\ 
    (1) There exists a local minimizer $\widehat{\bth}$ of the proposed objective function \eqref{FD_estimator} such that
    $$
\max_{k\in[K]}\|\widehat{\bth}_k-\bth^*_k\|_2\leq C\frac{\cG_{\max}}{\cG_{\min}}\sqrt{\frac{s_*\log M}{n}}\ \text{ and }\  \max_{k\in[K]}\|\widehat{\bth}_k-\bth^*_k\|_1\leq C \frac{\cG_{\max}^2}{\cG_{\min}^2}\sqrt{\frac{s_*^2\log M}{n}}
$$
for some constant $C > 0$, with probability approaching one as $n\rightarrow\infty$.\\
(2) Denote $\{\widehat{\cG}^j_i:i\in [\widehat{M}_j],\ j\in [p]\}$ as the group structure induced by $\widehat{\bth}$. Up to a permutation of group labels in $\big\{\widehat{\cG}^j_i: i\in [\widehat{M}_j];\ j\in [p]\big\}$, we have 
$$
\widehat{M}_j=M_j, \ \widehat{\cG}^j_i=\cG^j_i, \text{ for } i\in [\widehat{M}_j];\ j\in[p],
$$
with probability approaching one as $n\rightarrow\infty$.
\end{theorem}

Theorem~\ref{thm: rate of theta}(1) implies that the estimation error depends on $\cG_{\max}/\cG_{\min}$, reflecting imbalance in group-wise effective sizes. When the group sizes are balanced, so that $\cG_{\max}\approx \cG_{\min}$, the convergence rate reduces to the standard high-dimensional rate. Compared with the single-client estimator, whose $\ell_2$-error is $O_p\big(\sqrt{s_k \log p / n_k}\big)$ with $s_k$ denoting the sparsity of $\bth_k^*$ \citep{sasaki2024high}, the proposed PFE estimator improves accuracy when $n/n_k>(\cG_{\max}/\cG_{\min})^2(s_*/s_k)$ ignoring logarithmic factors, namely when the effective sample-size gain from federation exceeds the relative sparsity increase, after accounting for $(\cG_{\max}/\cG_{\min})^2$. 

Since the likelihood is based on a Pareto-type tail approximation rather than an exact parametric model, its stochastic and approximation errors must be controlled jointly. To this end, we use concentration inequalities and empirical process techniques to characterize the first- and second-order behavior of the tail-index regression loss, thereby controlling the approximation error. The development of such theoretical tools is of independent interest in extreme value theory.
The following corollary illustrates the special homogeneous case, where this improvement can be quantified explicitly.

\begin{corollary}[Convergence rate and group structure consistency of $\widehat{\bth}$ under homogeneous model]
\label{corollary: homogeneous rate of theta}
Suppose that the assumptions of Theorem \ref{thm: oracle property} hold. 
If all clients share the same regression coefficients, i.e., $\bth_1^*=\bth_2^*=\cdots=\bth_K^*$, then there exists a local minimizer $\widehat{\bth}$ of the proposed objective function \eqref{FD_estimator} such that $\widehat{\bth}_1=\widehat{\bth}_2=\cdots=\widehat{\bth}_K$,
$$
\max_k\|\widehat{\bth}_k-\bth^*_k\|_2 \le C \sqrt{\frac{s_1\log p}{n}}\ \text{ and }\ 
\max_k\|\widehat{\bth}_k-\bth^*_k\|_1 \le C \sqrt{\frac{s_1^2\log p}{n}}
$$
with probability approaching one as $n \to \infty$.
\end{corollary}

Corollary \ref{corollary: homogeneous rate of theta} shows that the proposed PFE estimator converges faster than the single-client estimator, with $\ell_2$-error improving from $O_p\big(\sqrt{s_1 \log p / n_k}\big)$ to $O_p\big(\sqrt{s_1 \log p / n}\big)$. This gain stems from aggregating information across all clients, reflecting the maximal improvement under homogeneous conditions.

\subsection{Federated estimation algorithm}

Solving \eqref{FD_estimator} is challenging because of the nonconcave sparsity and fusion penalties, while the empirical loss
$\cL_n(\bth) = (1/n)\sum_{k=1}^K \sum_{i=1}^{n_k} \ell(y_i^{(k)},\bx_i^{(k)},\boldsymbol{\theta}_k)$
has a non-globally Lipschitz gradient due to its exponential terms. Although the linearized ADMM algorithm of \citet{liu2025robust} is closely related and can handle nonconcave fusion penalties, its convergence theory requires the loss function to have a globally Lipschitz gradient or, equivalently for twice-differentiable functions, a bounded Hessian \citep{nishihara2015general,liu2019linearized,wang2019global,barber2024convergence}, which fails in our setting. 

To address this issue, we develop an ADMM-based algorithm with adaptive gradient descent \citep{berahas2024non}. The adaptive gradient step stabilizes the updates without global Lipschitz continuity, while the ADMM framework decomposes the optimization problem \eqref{FD_estimator} into tractable subproblems that can be solved distributively, making the method well suited for federated settings. We next describe the variable splitting scheme, iterative updates, federated implementation, and convergence guarantees of the proposed algorithm.

\noindent\textbf{ADMM algorithm.}  
To facilitate the ADMM formulation, let $\mathbf{E} = (\boldsymbol{e}_{i,K}-\boldsymbol{e}_{j,K}, \, i<j)^\T \in \mathbb{R}^{K(K-1)/2 \times K}$ and define $\mathbf{\Omega}_0 = \mathbf{E} \otimes \mathbf{I}_{p \times p} \in \mathbb{R}^{pK(K-1)/2 \times pK}$, where $\otimes$ denotes the Kronecker product. Let 
$\boldsymbol{\Omega}_1 = (\mathbf{I}_{pK\times pK}, \boldsymbol{\Omega}_0^\T)^\T \in \mathbb{R}^{(pK + pK(K-1)/2) \times pK}$.  
Introducing the auxiliary variable $\boldsymbol{\Delta} = \boldsymbol{\Omega}_1 \bth$, 
we rewrite \eqref{FD_estimator} as
\[
\min_{\bth, \boldsymbol{\Delta}} \cL_n(\bth) + h_\lambda(\boldsymbol{\Delta}) 
\quad \text{subject to} \quad \boldsymbol{\Omega}_1 \bth = \boldsymbol{\Delta},
\]
where 
$h_\lambda(\boldsymbol{\Delta}) = (1/K) \sum_{j=1}^{pK} p_{\lambda_1}(|\Delta_j|) 
+ (1/K) \sum_{j=pK+1}^{pK+pK(K-1)/2} p_{\lambda_2}(|\Delta_j|)$, 
and $\Delta_j$ is the $j$-th component of $\boldsymbol{\Delta}$.
The augmented Lagrangian is
\begin{align}\label{alg: augmented lagrangian}
    \mathcal{Q}_{\rho}(\bth, \boldsymbol{\Delta}, \boldsymbol{\zeta})
    = \cL_{n}(\bth)+ h_{\lambda}(\boldsymbol{\Delta}) 
    + \langle\boldsymbol{\zeta}, \boldsymbol{\Omega}_1 \bth-\boldsymbol{\Delta}\rangle
    + \frac{\rho}{2}\|\boldsymbol{\Omega}_1 \bth-\boldsymbol{\Delta}\|_2^2,
\end{align}
where $\boldsymbol{\zeta}$ is the Lagrange multiplier and $\rho$ is the penalty parameter. 
At iteration $t$, we update $\bth$, $\boldsymbol{\Delta}$, and $\boldsymbol{\zeta}$ sequentially by
\begin{align}\label{alg: zeta update}
& \bth^{(t+1)} = \underset{\bth}{\arg \min}\ \mathcal{Q}_{\rho^{(t)}}(\bth, \boldsymbol{\Delta}^{(t)}, \boldsymbol{\zeta}^{(t)}), \nonumber\\
& \boldsymbol{\Delta}^{(t+1)} = \underset{\boldsymbol{\Delta}}{\arg \min}\ \mathcal{Q}_{\rho^{(t)}}(\bth^{(t+1)}, \boldsymbol{\Delta}, \boldsymbol{\zeta}^{(t)}), \nonumber\\
& \boldsymbol{\zeta}^{(t+1)} = \boldsymbol{\zeta}^{(t)} + \rho^{(t)} \big( \boldsymbol{\Omega}_1 \bth^{(t+1)} - \boldsymbol{\Delta}^{(t+1)} \big),\ \rho^{(t+1)}=\sigma\rho^{(t)},\ \sigma>1.
\end{align}

\noindent\textbf{Update for $\bth^{(t+1)}$.}
The $\bth^{(t+1)}$ update minimizes a smooth loss with a quadratic penalty term and does not admit a closed-form solution. Moreover, since the gradient of the loss function is not globally Lipschitz continuous, standard descent arguments do not apply \citep{hong2016convergence,you2019nonconvex,liu2025robust}. We therefore use gradient descent with an adaptive stepsize based on a local curvature bound \citep{berahas2024non}.

Denote $f_t(\bth)=\mathcal{Q}_{\rho^{(t)}}(\bth, \boldsymbol{\Delta}^{(t)}, \boldsymbol{\zeta}^{(t)})$ and
$\overline{\boldsymbol{\Sigma}}_{k} = (1/n_{k}) \sum_{i=1}^{n_k}  \log(y_i^{(k)}/w_k)\bx_i^{(k)}\bx_i^{(k)\T}.$
Let $\bth^{(t)}=(\bth_1^{^{(t)}\T},\bth_2^{^{(t)}\T},\dots,\bth_K^{^{(t)}\T})^{\T}\in \mathbb{R}^{pK}$, where ${\bth_k^{(t)}}$ denotes the estimated regression coefficients for the $k$-th client at iteration $t$.
For any $\bth^{(t)}$ and radius $R>0$, define 
\begin{align}\label{alg: mut}
    \mu(\bth^{(t)},R)=\max_{k\in[K]}\big\{(n_k/n)\exp(\varpi_k(\bth_k^{(t)},R))\lambda_{\max}(\overline{\boldsymbol{\Sigma}}_{k})\big\}+\rho^{(t)}(K+1),
\end{align}
where 
$\varpi_k(\bth_k^{(t)},R)=\max_{i\in[n_k]}\big\{\bx_i^{(k)\T}\bth_k^{(t)}+
R\|\bx_i^{(k)}\|_2\big\}$.
Then $\mu(\bth^{(t)},R)$ upper bounds the largest eigenvalue of $\nabla^2 f_t(\bth)$ over $\|\bth-\bth^{(t)}\|_2\leq R$.
Using this local curvature bound, for step size factor $\eta>0$ and radius $R_t>0$, we update $\bth^{(t+1)}$ by
\begin{equation}\label{alg: theta update}
\bth^{(t+1)}=\bth^{(t)}-\frac{\eta}{\mu(\bth^{(t)},\widetilde{R}_t)}\nabla f_t(\bth^{(t)}),  
\end{equation}
where the gradient function $\nabla f_t(\bth^{(t)})$ and radius $\widetilde{R}_t$ are given by
\begin{align}
    &\nabla f_t(\bth^{(t)})=\nabla\cL_n(\bth^{(t)})+\boldsymbol{\Omega}_1^{\T}\boldsymbol{\zeta}^{(t)}+\rho^{(t)}\boldsymbol{\Omega}_1^\T(\boldsymbol{\Omega}_1 \bth^{(t)}-\boldsymbol{\Delta}^{(t)}),\label{alg: gradient ft}\\
    &\widetilde{R}_t=\max\big\{R_t,\eta\|\nabla f_t(\bth^{(t)})\|_2/\mu(\bth^{(t)},R_t)\big\}.\label{alg: tilde Rt}
\end{align}
The adaptive stepsize is constructed from a local Hessian bound and ensures stable descent without requiring global Lipschitz continuity of the gradient. Under suitable conditions, it guarantees $f_t(\bth^{(t+1)})<f_t(\bth^{(t)})$, which is key to convergence of the proposed algorithm.

\noindent\textbf{Update for $\boldsymbol{\Delta}^{(t+1)}$.}
The $\boldsymbol{\Delta}^{(t+1)}$ update corresponds to a proximal mapping associated with the nonconcave penalty:
$$
\begin{aligned}
\boldsymbol{\Delta}^{(t+1)} &= \underset{\boldsymbol{\Delta}}{\arg\min}\ h_{\lambda}(\boldsymbol{\Delta}) + 
\langle\boldsymbol{\zeta}^{(t)}, \boldsymbol{\Omega}_1 \bth^{(t+1)}-\boldsymbol{\Delta}\rangle+(\rho^{(t)}/2)\|\boldsymbol{\Omega}_1 \bth^{(t+1)}-\boldsymbol{\Delta}\|_2^2 \\
&=\underset{\boldsymbol{\Delta}}{\arg\min}\ (\rho^{(t)}/2)\|\boldsymbol{\Omega}_1\bth^{(t+1)}+ \boldsymbol{\zeta}^{(t)}/\rho^{(t)} -\boldsymbol{\Delta}\|_2^2+ h_{\lambda}(\boldsymbol{\Delta}).
\end{aligned}
$$
This optimization is separable across coordinates and admits closed-form thresholding updates. Let
$S_{\lambda}(z)= \operatorname{sgn}(z)(|z|-\lambda) \mathbbm{1}(|z|-\lambda>0)$
denote the soft-thresholding operator, and define
$T_{\lambda, \rho}(x)=\underset{z \in \mathbb{R}}{\arg \min} \big\{p_\lambda(z)+(\rho / 2)(z-x)^2\big\}.$
For the SCAD penalty, $T_{\lambda,\rho}^{\mathrm{SCAD}}(x)$ equals $S_{\lambda/\rho}(x)$ when $|x|\leq \lambda+\lambda/\rho$, equals $(a\rho-\rho)S_{a\lambda/(a\rho-\rho)}(x)/(a\rho-\rho-1)$ when $\lambda+\lambda/\rho<|x|\leq a\lambda$, and equals $x$ when $|x|>a\lambda$.
For the MCP penalty, $T_{\lambda, \rho}^{\text{MCP}}(x)$ is $\{a \rho/(a \rho-1)\} S_{\lambda / \rho}(x)$ when $|x| \leq a\lambda $, and equals $x$ when $|x|>a\lambda$.
Thus, the update is given componentwise by
\begin{align}\label{alg: delta update}
    \Delta^{(t+1)}_j=T_{\lambda,K\rho^{(t)}}\Big[\big(\boldsymbol{\Omega}_1\bth^{(t+1)}+ \boldsymbol{\zeta}^{(t)}/\rho^{(t)} \big)_j\Big], \text{ for } j\in [K(K-1)p/2+pK],
\end{align}
where $\big(\boldsymbol{\Omega}_1\bth^{(t+1)}+ \boldsymbol{\zeta}^{(t)}/\rho^{(t)} \big)_j$ is the $j$-th component of $\boldsymbol{\Omega}_1\bth^{(t+1)}+ \boldsymbol{\zeta}^{(t)}/\rho^{(t)}$, with $\lambda=\lambda_1$ for $j\in[pK]$ and $\lambda=\lambda_2$ otherwise. The closed-form update makes the $\boldsymbol{\Delta}^{(t+1)}$ step computationally efficient despite the nonconcave penalties.

\begin{algorithm}[!tb]
\caption{Personalized federated estimation (PFE) algorithm}
\label{alg: fed estimation}
\KwInput{ 
Observations $\mathcal{D}_k = \big\{\bx_{i}^{(k)},y_{i}^{(k)}\big\}_{i=1}^{N_k}$, $k=1,\dots,K$; Parameters $\lambda_1, \lambda_2$, $a$, $\rho^{(0)}$, $\sigma$, $\eta$, $\{R_t\}_{t=1}^{\infty}$ and
number of iterations $T$.}
{Initialization}: Set $\bth^{(0)}=\boldsymbol{\Delta}^{(0)}=\boldsymbol{\zeta}^{(0)}=\boldsymbol{0}$. Each client sends $n_k$ and 
$\lambda_{\max}(\overline{\boldsymbol{\Sigma}}_{k})$ to the server;

{\bf{for} $t=0,1,\dots,T-1$ \bf{do}}

\quad \textbf{Clients $k\in[K]$ do in parallel:}

\quad  \quad Download $\bth_k^{(t)}$ from server, compute $\nabla\cL_{n,k}(\bth_k^{(t)})= \sum_{i=1}^{n_k} 
\nabla \ell\big(y_{i}^{(k)}, \bx_{i}^{(k)}, \bth_k^{(t)}\big)$;

\quad  \quad Calculate $\big\{\bx_i^{(k)\T}\bth_k^{(t)}\big\}_{i\in[n_k]}$ and $\{\|\bx_i^{(k)}\|_2\}_{i\in[n_k]}$;

\quad \quad Upload $\nabla\cL_{n,k}(\bth_k^{(t)})$, $\big\{\bx_i^{(k)\T}\bth_k^{(t)}\big\}_{i\in[n_k]}$ and $\{\|\bx_i^{(k)}\|_2\}_{i\in[n_k]}$ to server; 

\quad  \textbf{Server:}

\quad  \quad Calculate $\nabla f_t(\bth^{(t)})$ by \eqref{alg: gradient ft}, $\mu(\bth^{(t)},R_t)$ by \eqref{alg: mut};

\quad  \quad Calculate $\widetilde{R}_t$ by \eqref{alg: tilde Rt}, and $\mu(\bth^{(t)}, \widetilde{R}_t)$ by \eqref{alg: mut};

\quad  \quad Update $\bth^{(t+1)}$ by \eqref{alg: theta update};

\quad \quad Update $\boldsymbol{\Delta}^{(t+1)}$ by \eqref{alg: delta update} and $\boldsymbol{\zeta}^{(t+1)}$ by \eqref{alg: zeta update};

{\bf{end for}}

\KwOutput{Client $k\in[K]$ return $\widehat{\bth}_k=\bth_k^{(T)}$ and server return $\widehat{\bth}=\bth^{(T)}$.}
\end{algorithm}

\noindent\textbf{Federated implementation.} 
Initially, each client sends $n_k$ and 
$\lambda_{\max}(\overline{\boldsymbol{\Sigma}}_{k})$ to the server. The update of $\bth^{(t+1)}$ requires the gradient $\nabla \cL_n(\bth^{(t)})$, which depends on local client data. At iteration $t$, client $k$ receives $\bth_k^{(t)}$, computes the local gradient
$\nabla \cL_{n,k}(\bth^{(t)}_k) = \sum_{i=1}^{n_k} 
\nabla \ell\big(y_{i}^{(k)}, \bx_{i}^{(k)}, \bth^{(t)}_k\big)$ together with $\big\{\bx_i^{(k)\T}\bth_k^{(t)}\big\}_{i\in[n_k]}$ and $\{\|\bx_i^{(k)}\|_2\}_{i\in[n_k]}$, and uploads these quantities to the server. The server aggregates the local gradients to form $\nabla \cL_n(\bth^{(t)})$, computes 
$\nabla f_t(\bth^{(t)})$, $\mu(\bth^{(t)},R_t)$, $\widetilde{R}_t$, and $\mu(\bth^{(t)}, \widetilde{R}_t)$, and updates $\bth^{(t+1)}$.
The updates of $\boldsymbol{\Delta}^{(t+1)}$ and $\boldsymbol{\zeta}^{(t+1)}$ do not involve raw client data and are computed entirely on the server. The procedure is iterated until convergence. Algorithm~\ref{alg: fed estimation} summarizes the proposed federated algorithm. The method preserves data privacy by exchanging only gradients and summary statistics. 

Let $\{\bth^{(t)},\boldsymbol{\Delta}^{(t)},\boldsymbol{\zeta}^{(t)}\}_{t=1}^{\infty}$ be the sequence generated by Algorithm~\ref{alg: fed estimation}, with $\bth^{(0)}=\boldsymbol{\Delta}^{(0)}=\boldsymbol{\zeta}^{(0)}=\boldsymbol{0}$. We have the following convergence result of the proposed  algorithm.

\begin{proposition}[Algorithm convergence]\label{prop: alg convergence}
    Suppose that Assumption \ref{assumption: penalty conditions}  hold. If $\eta\in(0,2)$, $\{R_t\}_{t=1}^\infty$ is bounded, $\rho^{(0)}>2v/K$, and $g\big(a,\min\{\lambda_1,\lambda_2\}\big)/K>\cL_{n}(\boldsymbol{0})-\min_{\bth}\cL_{n}(\bth)+\big\{c_\lambda\sigma(\sigma+1)\big\}/\big\{2\rho^{(0)}(\sigma-1)\big\}$, where $c_\lambda>0$ is a constant depending on $\lambda_1$ and $\lambda_2$. 
    Then, the sequence $\big\{\bth^{(t)},\boldsymbol{\Delta}^{(t)},\boldsymbol{\zeta}^{(t)}\big\}_{t=1}^{\infty}$ is bounded and $\lim_{T\to \infty}\sum_{t=0}^T\|\bth^{(t+1)}-\bth^{(t)}\|_2^2<\infty$. Moreover, there exists a subsequence $\big\{\bth^{(t_k)},\boldsymbol{\Delta}^{(t_k)},\boldsymbol{\zeta}^{(t_k)}\big\}_{t_k=1}^{\infty}$ that converges to a stationary point satisfying the KKT condition of the augmented Lagrangian \eqref{alg: augmented lagrangian}.
\end{proposition}

Proposition \ref{prop: alg convergence} shows that the proposed ADMM algorithm with adaptive gradient updates generates a bounded sequence with vanishing successive differences and converges to a stationary point. The result establishes convergence despite the non-Lipschitz gradient of the loss function and the nonconcave penalty.

\section{Personalized federated inference}\label{sec:FD_inference}

\subsection{Methods and theoretical properties}

In this section, we develop personalized federated inference (PFI) procedures for the PFE estimator $\widehat{\bth}$. Existing debiased methods are typically constructed from a single dataset \citep{sasaki2024high}, and therefore cannot exploit structural similarities across clients. 
A natural alternative is to perform inference based on the oracle property of sparse estimators, where one first identifies the support set of the regression coefficients $\bbeta^*$, and then conducts classical low-dimensional inference for $\widehat{\bbeta}$ as if the true support were known in advance. Theorem \ref{thm: oracle property} guarantees that the resulting estimator $\widehat{\boldsymbol{\theta}}$ is asymptotically equivalent to the oracle estimator $\mathbf{U}\widehat{\bbeta}$, thereby allowing low-dimensional inferential results to be transferred to the original high-dimensional problem.
However, such oracle-property-based inference usually relies on consistent variable selection, and hence requires beta-min conditions on all nonzero coefficients \citep{yang2019high}. Moreover, when the related clients have heterogeneous covariate distributions or different effective sample sizes, treating the combined as a pooled sample may be statistically inefficient \citep{gu2023distributed}.

To address this issue, we adopt a weighted debiasing strategy \citep{gu2023distributed}. The key idea is to first construct local debiased estimators on individual clients and then aggregate only those estimators corresponding to clients that share the same coefficient. This preserves the client-specific interpretation of the target parameter while allowing the aggregation weights to adapt to client-level heterogeneity.

We first consider the case where the group structures $\{\mathcal{G}_i^j:i\in[M_j],j\in[p]\}$ are known; when they are unknown, we use the estimated groups from Section~\ref{sec:FD_problem}. 
For the $j$-th coefficient $\theta_{\ell j}^*$ on client $\ell$, suppose that $\ell\in\mathcal{G}_i^j$. Since $\theta_{\ell j}^*$ is common to all clients in $\mathcal{G}_i^j$, we construct its debiased estimator $\widehat{\theta}_{\ell j}^{de}$ by aggregating the local debiased estimators from the related clients $\{\cD_k:k\in\mathcal{G}_i^j\}$ rather than only $\cD_\ell$. Specifically,
\begin{equation}\label{weighted debiased estimator}
\widehat{\theta}_{\ell j}^{de} = \sum_{k \in \mathcal{G}_i^j} v_k \widetilde{\theta}_{k j}^{de}, \quad \text{for } \ell \in \mathcal{G}_i^j,
\end{equation}
where $\widetilde{\theta}_{kj}^{de}$ is the local debiased estimator computed on client $k$, and $\{v_k:k\in\mathcal{G}_i^j\}$ are aggregation weights with $\sum_{k\in\mathcal{G}_i^j}v_k=1$ and $v_k>0$.
These weights will be chosen later to minimize the asymptotic variance of $\widehat{\theta}_{\ell j}^{de}$, so that the resulting estimator exploits the shared coefficient structure while accounting for heterogeneity in local sample sizes and covariate distributions.

Define the local debiased estimator $\widetilde{\theta}_{kj}^{de}$ for the $j$-th coefficient on client $k$ as
\begin{equation}\label{local debias}
    \widetilde{\theta}_{k j}^{de}= \widehat{\theta}_{k j} - \widehat{\boldsymbol{u}}_{j}^{(k)\T}\frac{1}{n_{k}}\sum_{i=1}^{n_k} \{\exp{(\bx_i^{(k)\T}\widehat{\bth}_{k})\log(y_i^{(k)}/w_k)-1}\}\bx_i^{(k)},
\end{equation}
where $\widehat{\bth}_{k}=(\widehat{\theta}_{k1},\ldots,\widehat{\theta}_{kp})^{\T}$ is the client-specific PFE estimator obtained from Algorithm~\ref{alg: fed estimation}. 
Let $\widehat{\boldsymbol{\Sigma}}_{k}=n_k^{-1}\sum_{i=1}^{n_k}\bx_i^{(k)}\bx_i^{(k)\T}$. The projection direction $\widehat{\boldsymbol{u}}_{j}^{(k)}$ is obtained by solving the quadratic programming problem
\begin{equation}\label{projection direction}
\begin{aligned}
&\widehat{\boldsymbol{u}}_{j}^{(k)}=  \underset{\|\boldsymbol{u}\|_1\leq C}{\arg\min}\  \boldsymbol{u}^{\top}\widehat{\boldsymbol{\Sigma}}_{k} \boldsymbol{u} \\
\text { s.t. }  &\big\|\widehat{\boldsymbol{\Sigma}}_{k} \boldsymbol{u}-\boldsymbol{e}_{j,p}\big\|_{\infty} \leq \mu_{k}, \text{ and } \max_{i\in [n_k]}|\bx_i^{(k)\T}\boldsymbol{u}|\leq \gamma_{k},
\end{aligned}
\end{equation}
where $C$ is a large positive constant, $\boldsymbol{e}_{j,p}$ denotes the $j$-th canonical basis of the Euclidean space $\mathbb{R}^p$, $\mu_{k}=C'\sqrt{(\log p)/n_{k}}$ and $\gamma_{k}=C''\sqrt{\log n_{k}}$ for some constants $C',C''>0$. 
The following theorem establishes the asymptotic normality of the local debiased estimator.

\begin{theorem}(Local debiased estimator)\label{thm: local debiased estimator}
    Suppose that the assumptions of Theorem \ref{thm: oracle property} hold. Assume further that 
    $\max_{j\in[p]}\|(\boldsymbol{\Sigma}^{(k)}_{w_k})^{-1}\boldsymbol{e}_{j,p}\|_1\leq C$ for some constant $C>0$, and $\sqrt{n_k}w_k^{-\underline{\beta}}\rightarrow 0$ as $n_k\rightarrow \infty$.  If $n\gg (\cG_{\max}^4/\cG_{\min}^4)s_*^2(\log M)\sqrt{n_k}$,
    then 
    $$
    \widetilde{V}_{k j}^{-1/2}\big(\widetilde{\theta}_{kj}^{de}-\theta^*_{kj}\big)\overset{d}{\rightarrow} \mathcal{N}(0,1)
    $$ 
    as $(n_k, p)\to \infty$, where $\widetilde{V}_{k j} = \widehat{\boldsymbol{u}}_{j}^{(k)\T} \widehat{\boldsymbol{\Sigma}}_{k} \widehat{\boldsymbol{u}}_{j}^{(k)}/n_k$ is the asymptotic variance estimator of $\widetilde{\theta}_{kj}^{de}$.
\end{theorem}

Theorem~\ref{thm: local debiased estimator} shows that each local debiased estimator is asymptotically normal with variance $\widetilde{V}_{kj}$. Unlike the debiased procedure in \cite{sasaki2024high}, which relies on sample splitting and cross-fitting, the proposed construction uses all local tail observations in the debiasing step and avoids the additional randomness introduced by data splitting.

Leveraging the asymptotic normality of the local debiased estimators in Theorem~\ref{thm: local debiased estimator}, the asymptotic variance of the aggregated estimator $\widehat{\theta}_{\ell j}^{de}$ is $\widehat{V}_{\ell j} = \sum_{k\in\cG_i^j} v_k^2 \widetilde{V}_{k j}$. 
To achieve the highest efficiency, we choose the aggregation weights by minimizing $\widehat{V}_{\ell j}$ subject to $\sum_{k\in\cG_i^j}v_k=1$ and $v_k>0$, leading to the convex optimization problem 
$$
\underset{\{v_{k}:k\in\cG_i^j\}}{\arg\min} \sum_{k\in\cG_i^j} v_{k}^2\widetilde{V}_{k j},\; \text{subject to } \sum_{k\in\cG_i^j}v_{k}=1, v_k>0.
$$ 
The optimal weights are $v_k = \widetilde{V}_{k j}^{-1} / \sum_{k\in\cG_i^j} \widetilde{V}_{k j}^{-1}$, which assigns larger weight to clients with smaller local variance. The following theorem establishes the asymptotic normality of the weighted debiased estimator.

\begin{theorem}[Weighted debiased estimator with known group structure]\label{thm: weighted debiased estimator}
    Suppose that the assumptions of Theorem \ref{thm: oracle property} hold. Assume further that  
    $\max_{j\in[p]}\|(\boldsymbol{\Sigma}^{(k)}_{w_k})^{-1}\boldsymbol{e}_{j,p}\|_1\leq C$ and $\max_{i\in[M_j],j\in[p]}|\cG_i^j|\leq C$ for some constant $C>0$, and $\sqrt{n_{\max}^{\operatorname{client}}}w_{\min}^{-\underline{\beta}}\rightarrow 0$ as $n_{\min}^{\operatorname{client}}\rightarrow \infty$.  If $n\gg (\cG_{\max}^4/\cG_{\min}^4)s_*^2(\log M)\sqrt{n_{\max}^{\operatorname{client}}}$ and $v_k = \widetilde{V}_{k j}^{-1} / \sum_{k\in\cG_i^j} \widetilde{V}_{k j}^{-1}$,
     then for any $i\in [M_j]$, $j\in [p]$, and any $\ell\in\cG_i^j$, we have
    $$
    \widehat{V}_{\ell j}^{-1/2}\big(\widehat{\theta}_{\ell j}^{de}-\theta^*_{\ell j}\big)\overset{d}{\rightarrow} \mathcal{N}(0,1), 
    $$
    as $(n_{\min}^{\operatorname{client}}, p)\to \infty$, where $\widehat{V}_{\ell j}=1/\sum_{k\in\cG_i^j} \widetilde{V}_{k j}^{-1}$ is the asymptotic variance estimator of $\widehat{\theta}_{\ell j}^{de}$ with optimal weights $v_k$.
\end{theorem}

Let $z_{\alpha}$ denotes the upper $\alpha$ quantile of $\mathcal{N}(0,1)$. The asymptotic $1-\alpha$ confidence interval of $\theta^*_{\ell j}$ based on the weighted debiased estimator is given by
\[
\mathrm{CI}_{1-\alpha}(\theta^*_{\ell j}) 
= \left[\, 
\widehat{\theta}_{\ell j}^{de} - z_{\alpha/2}\, \widehat{V}_{\ell j}^{1/2}, \;\; 
\widehat{\theta}_{\ell j}^{de} + z_{\alpha/2}\, \widehat{V}_{\ell j}^{1/2} 
\,\right].
\]

Theorem~\ref{thm: weighted debiased estimator} ensures that the weighted debiased estimator remains asymptotically normal and achieves a variance that is strictly smaller than any individual local estimator when $|\cG_i^j|\ge 2$, i.e., $\widehat{V}_{\ell j}<\widetilde{V}_{k j}$ for any $\ell, k \in \cG_i^j$. Specifically, denote the maximal local variance $V_{i,\max}^j=\max_{k\in\cG_i^j}\widetilde{V}_{k j}$ and the minimal local variance $V_{i,\min}^j=\min_{k\in\cG_i^j}\widetilde{V}_{k j}$, then $V_{i,\min}^j/|\cG_i^j|\leq \widehat{V}_{\ell j}\leq  V_{i,\max}^j/|\cG_i^j|$. If the local variances are approximately equal, i.e., $V_{i,\max}^j \approx V_{i,\min}^j$, the asymptotic variance is reduced roughly by a factor of $1/|\cG_i^j|$, and correspondingly, the confidence interval is shorter than that of any local debiased estimator by a factor of approximately $1/\sqrt{|\cG_i^j|}$.
Importantly, this efficiency gain does not require the local datasets to follow identical distributions, allowing for heterogeneity across data sources.

Building on Theorem \ref{thm: weighted debiased estimator}, we extend the results to the case of unknown group structures.

\begin{corollary}[Weighted debiased estimator with unknown group structure]
    Suppose that the assumptions of Theorem~\ref{thm: weighted debiased estimator} hold. If the estimated groups $\{\widehat{\cG}_i^j:i\in[\widehat M_j],j\in[p]\}$ from Theorem~\ref{thm: rate of theta}(2) are used in the definitions of $\widehat{\theta}_{\ell j}^{de}$ and $\widehat{V}_{\ell j}$, then the conclusion of Theorem~\ref{thm: weighted debiased estimator} continues to hold.
\end{corollary}

\subsection{Federated inference algorithm}

The proposed PFI procedure can be implemented with two rounds of communication. In the first round, each client computes its local debiased estimators and variance estimators, $\{(\widetilde{\theta}_{kj}^{de},\widetilde{V}_{kj})\}_{k\in[K],j\in[p]}$, and sends them to the server. In the second round, the server aggregates the received estimators within each estimated group using the inverse-variance weights, and returns the resulting weighted debiased estimators $\widehat{\theta}_{\ell j}^{de}$ and variance estimators $\widehat{V}_{\ell j}$ to the corresponding clients. The overall PFI procedure is summarized in Algorithm \ref{alg: fed inference}.

\begin{algorithm}[!tb]
\caption{Personalized federated inference (PFI) algorithm}
\label{alg: fed inference}
\KwInput{ 
Observations $\mathcal{D}_k = \big\{\bx_{i}^{(k)},y_{i}^{(k)}\big\}_{i=1}^{N_k}$, $k\in[K]$; Estimator $\big\{\widehat{\bth}_k\big\}_{k\in[K]}$ from Algorithm~\ref{alg: fed estimation} and $\big\{\widehat{\cG}_i^j\big\}_{i\in [\widehat{M}_j],j\in[p]}$; Parameters $\mu_k$ and $\gamma_k$.}

\textbf{Clients $k\in[K]$ do in parallel:}

\quad Calculate $\widehat{\boldsymbol{u}}_{j}^{(k)}$ by \eqref{projection direction}, $\widetilde{\theta}_{kj}^{de}$ by \eqref{local debias}, and $\widetilde{V}_{k j} = \widehat{\boldsymbol{u}}_{j}^{(k)\T} \widehat{\boldsymbol{\Sigma}}_{k} \widehat{\boldsymbol{u}}_{j}^{(k)}/n_k
$ for $j\in[p]$;

\quad Upload $\big\{\big(\widetilde{\theta}_{kj}^{de},\widetilde{V}_{k j}\big)\big\}_{j\in[p]}$ to server;
 
\textbf{Server:}

\quad For $\ell\in \widehat{\cG}_i^j, j\in[p]$, calculate $v_{k} = \widetilde{V}_{k j}^{-1}/\big(\sum_{k\in\widehat{\cG}_i^j}\widetilde{V}_{k j}^{-1}\big)$ and $\widehat{V}_{\ell j} =\sum_{k\in\widehat{\cG}_i^j} v_{k}^2\widetilde{V}_{k j}$;

\quad Compute $\widehat{\theta}_{\ell j}^{de}$ by \eqref{weighted debiased estimator} with $\cG_i^j$ replaced by $\widehat{\cG}_i^j$;

\KwOutput{Server return $\big(\widehat{\theta}_{\ell j}^{de},\widehat{V}_{\ell j}\big)$ for $\ell\in \widehat{\cG}_i^j, j\in[p]$}
\end{algorithm}

\section{Simulation studies}\label{sec:simu}

We evaluate the finite-sample performance of the proposed methods through simulation studies. All results are based on 500 replications. We consider both heterogeneous and homogeneous coefficient structures: (i) Heterogeneous: the clients are divided into two equally sized groups, with $\bth_k^* = (2, -2, -2, -2, \mathbf{0}_{p-4}^{\top})^{\top}$ for $k\in\{1, \ldots, K / 2\}$, and $\bth_k^* = (-2, 2, 2, -2, \mathbf{0}_{p-4}^{\top})^{\top}$ for $k\in\{K/2+1, \ldots, K\}$;
(ii) Homogeneous: all clients share the same coefficient $\bth_k^* = (2, -2, -2, -2, \mathbf{0}_{p-4}^{\top})^{\top}$, $k\in\{1, \ldots, K\}$.
In all settings, the clients have equal sample sizes, that is $N_1=\cdots=N_K$, and the data $\{(y_i^{(k)},\bx_i^{(k)})\}$ are generated independently according to the models below, and the conditional tail index is $\alpha_k(\bx^{(k)})=\exp\{\bx^{(k)\top}\bth_k^*\}$.

\noindent\textbf{Data generating process of covariates $\bx^{(k)}$:} \\
\textbf{(XI)} (Uniform design): 
    $x_{i j}^{(k)}=\sqrt{12}\left\{\Phi\left(z_{ij}\right)-1 / 2\right\}$ for $j=1, \ldots, p$, where $z_{ij}$ is a standard normal random variable with pairwise covariance $\operatorname{Cov}\left(z_{ij_1}, z_{ij_2}\right)=0.5^{\left|j_1-j_2\right|}$ for $1 \leq j_1, j_2 \leq p$. The marginal distributions of $x_{i j}^{(k)}$ are Uniform $[-\sqrt{3}, \sqrt{3}]$; \\
\textbf{(XII)} (Gaussian design): 
    $\bx_i^{(k)} \sim \mathcal{N}\left(\boldsymbol{0}, \boldsymbol{\Sigma}\right)$, where $\boldsymbol{\Sigma}=(0.5^{|j_1-j_2|})_{p\times p}$.

\noindent\textbf{Data generating process of response $y^{(k)}$:} \\
\textbf{(YI)}: The conditional tail distribution is 
    $
    \mathbb{P}(y^{(k)} > t \mid \bx^{(k)}) =\{(1+m) t^{-\alpha_k(\bx^{(k)})}\}/\{1+m t^{-\alpha_k(\bx^{(k)})}\},
    $
    where $m=0.3$; \\
\textbf{(YII)}: The distribution of $y^{(k)}$ conditional on $\bx^{(k)}$ is a $t$-distribution with degrees of freedom $\alpha_k(\bx^{(k)})$; \\
\textbf{(YIII)}: The distribution of $y^{(k)}$  conditional on $\bx^{(k)}$ is a Burr Type XII distribution with scale parameter $1$, and shape parameters $2$ and $\alpha_k(\bx^{(k)})/2$. The tail conditional distribution is given by
    $
    \mathbb{P}(y^{(k)} > t \mid \bx^{(k)})= \left(1 + t^2\right)^{-\alpha_k(\bx^{(k)})/2}.
    $ \\
\textbf{(YIV)}: The distribution of $y^{(k)}$  conditional on $\bx^{(k)}$ is a Fréchet distribution with location parameter 0, scale parameter $m$, and shape parameter $\alpha_k(\bx^{(k)})$. The tail conditional distribution is given by
    $
    \mathbb{P}(y^{(k)} > t \mid \bx^{(k)})= 1 - \exp \big\{-m^{\alpha(\bx^{(k)})} t^{-\alpha_k(\bx^{(k)})}\big\},
    $
    where $m = 1$.

\subsection{Personalized federated estimation}

In this subsection, we evaluate the proposed personalized federated estimator (PFE) for estimating tail index regression coefficients. We compare PFE with three benchmarks: (i) Oracle estimator (Oracle): The estimator obtained by \eqref{oracle estimator}, where the group structures are known.
(ii) Individual estimator (Indv) \citep{sasaki2024high}: The regularized estimator obtained locally on each client $k$ by
\begin{equation}\label{local estimator}
    \widehat{\bth}_k^{\text{Indv}}(w_k,\lambda) = \underset{\boldsymbol{\theta}_k \in \mathbb{R}^{p}}{\arg \min }\ \frac{1}{n_k} \sum_{i=1}^{n_k}  \ell\big(y_{i}^{(k)},\bx_{i}^{(k)},\boldsymbol{\theta}_k\big)+ \sum_{j=1}^{p}p_{\lambda}(|\theta_{kj}|).
\end{equation}
(iii) Federated Learning Averaged Estimator (FDAV): The estimator obtained by simply averaging the individual estimators from \eqref{local estimator} across all clients.

\subsubsection{Tuning parameter selection}\label{sec:tuning_select}

In the proposed personalized federated framework, the local thresholds $w_k$ determine the effective tail samples on each client, whereas the global regularization parameters $\lambda_1$ and $\lambda_2$ control sparsity and fusion in the federated estimator. We use a two-step tuning procedure: each client first selects $w_k$ by the discrepancy measure of \cite{wang2009tail}; given the selected thresholds, the server selects $(\lambda_1,\lambda_2)$ by Bayesian information criterion (BIC) \citep{ma2017concave,tang2016fused,liu2025robust}.

\noindent\textbf{Selection of $w_k$.}
The choice of $w_k$ involves the usual bias--variance trade-off in extreme value theory: a lower threshold yields more exceedances but may violate the Pareto-type approximation, whereas a higher threshold improves tail approximation but reduces the effective sample size. We use the discrepancy measure of \cite{wang2009tail}, which is based on the probability integral transform implied by the fitted tail model. Under the Pareto-type approximation,
$\bbP\big(y^{(k)}>t\mid y^{(k)}>w_k,\bx^{(k)}\big)\approx\exp\{-\alpha_k(\bx^{(k)})\log(t/w_k)\}$ for $t>w_k$. Hence, if the model is well fitted above $w_k$, 
$U_i^{(k)}=\exp\{-\alpha_k(\bx_i^{(k)})\log(y_i^{(k)}/w_k)\}$ should be approximately uniform on $[0,1]$. Replacing $\alpha_k(\bx_i^{(k)})$ by $\exp\{\bx_i^{(k)\T}\widehat{\bth}_k(w_k,\lambda)\}$ gives
$$\widehat{U}_i^{(k)}=\exp \Big\{-\exp \big(\bx_i^{(k)\top} \widehat{\bth}_k(w_k,\lambda)\big) \log (y_i^{(k)} / w_k)\Big\},$$
where $\widehat{\bth}_k(w_k,\lambda)$ is the local estimator in \eqref{local estimator}. Thus, a good threshold should make the empirical distribution of $\{\widehat U_i^{(k)}:y_i^{(k)}>w_k\}$ close to the uniform distribution.

Let $\widehat F_{n_k}$ be the empirical distribution function of $\{\widehat U_i^{(k)}:y_i^{(k)}>w_k\}$. We measure the departure from uniformity by
$
\widehat{D}\left(w_k,\lambda\right)=n_k^{-1} \sum_{i=1}^{n_k}\big\{\widehat{U}_i^{(k)}-\widehat{F}_{n_k}(\widehat{U}_i^{(k)})\big\}^2.
$ 
We then select
$$w_k^*=\operatorname{argmin}_{w_k, \lambda}\widehat{D}\left(w_k,\lambda\right).$$
In practice, we search $100$ values for $\lambda$ over $\big[0.5\sqrt{\log p/n_k},,5\sqrt{\log p/n_k}\big]$, and a grid of $100$ values for $w_k$ corresponding to sample fractions (i.e., $n_k/N_k$) evenly spaced in $(0.1,1)$.

\noindent\textbf{Selection of $\lambda_1$ and $\lambda_2$.}
Given the selected thresholds $w_k$, we choose $(\lambda_1,\lambda_2)$ by BIC \citep{ma2017concave,tang2016fused,liu2025robust}. 
In federated setting, each client $k$ transmits its local loss
$\cL_{n,k}(\widehat{\boldsymbol{\theta}}_k) = \sum_{i=1}^{n_k} 
\ell\big(y_{i}^{(k)}, \bx_{i}^{(k)}, \widehat{\boldsymbol{\theta}}_k\big)$ to the server, which computes  BIC as
$$\operatorname{BIC}\left(\lambda_1, \lambda_2\right)=\log \left\{\frac{1}{n} \sum_{k=1}^K  \sum_{i=1}^{n_k} \ell\big(y_{i}^{(k)},\bx_{i}^{(k)},\widehat{\boldsymbol{\theta}}_k\big)\right\}+ \frac{\log n}{n} \cdot \sum_{j=1}^p \widehat{K}_j\left(\lambda_1, \lambda_2\right),$$ 
where $\widehat{K}_j(\lambda_1, \lambda_2)$ denotes the number of distinct coefficients for variable $j$ in the estimated $\widehat{\bth}$. In practice, we consider a grid of 100 values for each of $\lambda_1$ and $\lambda_2$ over the interval $\big[0.5\sqrt{\log (pK)/n},5\sqrt{\log (pK)/n}\big]$ and select the pair $(\lambda_1^*, \lambda_2^*)$ that minimizes the BIC value. 
These tuning steps preserve the federated learning requirements, since only local computations and aggregated information are exchanged, keeping raw data private.

In Algorithm~\ref{alg: fed estimation}, we choose the parameters $\eta=0.5$, $\rho^{(0)}=0.2$, $\sigma=1.1$, $R_t=0.2$, and $a=5$ for both the SCAD and MCP penalties.

\subsubsection{Evaluation metrics and results}

We evaluate estimation performance using three metrics: 
(a) Averaged mean squared error: 
   $ \mathrm{AMSE}
    = \|\widehat{\bth}-\bth^*\|_2^2/K$;
(b) $F_1$-score $=2\left(\text {Recall}^{-1}+\text {Precision}^{-1}\right)^{-1}$, where
\begin{equation*}
    \mathrm{Recall}
    =
    \frac{\#\{j\in [pK],\widehat{\theta}_j\neq0, \theta^*_j\neq0 \}}{\#\{j\in [pK],\theta^*_j\neq0\}},
    \qquad
    \mathrm{Precision} 
    =
    \frac{\#\{j\in [pK],\widehat{\theta}_j\neq0, \theta^*_j\neq0 \}}{\#\{j\in [pK],\widehat{\theta}_j\neq0\}};
\end{equation*}
(c) $\mathrm{Recovery} = \sum_{j=1}^p \widehat{M}_j/ \sum_{j=1}^p M_j$, where $M_j$ is the true number of distinct coefficients for covariate $j$ in assumption \ref{assumption: parameter group structure}, and $\widehat{M}_j$ is the estimate.
This metric measures the recovery of the latent personalization structure. 

Table~\ref{simu_table_est} reports the results with the SCAD penalty, while the MCP results are given in Table~\ref{simu_table_est_MCP} of the supplementary material. Across both heterogeneous and homogeneous scenarios, PFE achieves AMSE close to the oracle estimator and substantially smaller than the individual estimator, showing the benefit of borrowing information across related clients. The comparison with FDAV further highlights the role of personalization: under heterogeneity, FDAV suffers from large AMSE because simple averaging ignores client-specific coefficient differences, whereas PFE maintains low AMSE and Recovery values close to one. This indicates that PFE can effectively exploit shared structure while preserving client-level heterogeneity.

\begin{table*}[!tb]
	\centering
    \caption{Simulation results for Model XI with $(K,p)=(10,50)$ under heterogeneous and homogeneous scenarios.}
	\begingroup
	\setlength{\tabcolsep}{2.6pt} 
	\renewcommand{\arraystretch}{0.8} 
        \label{simu_table_est}
	\begin{tabular}{cccccccccccc}
		\toprule
        &  && \multicolumn{3}{c}{Heterogeneous}  && \multicolumn{3}{c}{Homogeneous}   \\
        \cline{4-6} \cline{8-10}
		  Model & Method   &&  AMSE  & $F_1$-score & Recovery && AMSE  & $F_1$-score & Recovery \\ \hline
		\multirow{4}{*}{\makecell{Model YI \\ $N_k=400$, $n/N=0.625$}} 
		& PFE   && 0.114 & 0.999 & 1.083  && 0.079 & 0.999 & 1.028\\
        & Oracle  && 0.085 & 1.000 & 1.000 && 0.074 & 1.000 & 1.000  \\
		& Indv   && 0.228 & 0.999 & 1.623 && 0.236 & 1.000 & 1.721 \\
        & FDAV  && 12.028 & 0.998 & 0.943 && 0.078 & 1.000 & 1.000 \\\hline
        \multirow{4}{*}{\makecell{Model YII \\ $N_k=1600$, $n/N=0.144$}} 
		& PFE   && 0.055 & 1.000 & 1.004 && 0.042 & 1.000 & 1.004 \\
        & Oracle &&  0.050 & 1.000 & 1.000 && 0.040 & 1.000 & 1.000 \\
		& Indv   && 0.219 & 0.999 & 1.623 && 0.219 & 1.000 & 1.721 \\
        & FDAV  && 12.021 & 0.998 & 0.943 && 0.065 & 1.000 & 1.000 \\\hline
        \multirow{4}{*}{\makecell{Model YIII \\  $N_k=800$, $n/N=0.344$}} 
		& PFE   &&  0.055 & 0.999 & 1.028 && 0.018 & 1.000 & 1.001\\
        & Oracle  && 0.034 & 0.999 & 1.000 && 0.017 & 1.000 & 1.000  \\
		& Indv   && 0.177 & 1.000 & 1.623 && 0.175 & 0.999 & 1.720 \\
        & FDAV  && 12.017 & 0.997 & 0.943 && 0.021 & 1.000 & 1.000 \\\hline
         \multirow{4}{*}{\makecell{Model YIV \\ $N_k=1600$, $n/N=0.165$}} 
		& PFE   && 0.152 & 1.000 & 1.021 && 0.178 & 0.998 & 1.000 \\
        & Oracle  && 0.151 & 1.000 & 1.000 && 0.140 & 1.000 & 1.000  \\
		& Indv   &&  0.281 & 1.000 & 1.623 && 0.283 & 1.000 & 1.721 \\
        & FDAV  && 12.045 & 1.000 & 0.943 && 0.191 & 0.998 & 1.000 \\
		\bottomrule
	\end{tabular}
    \endgroup
\end{table*}

\begin{figure}[!tb]
\centering
\includegraphics[width=0.98\textwidth, height=0.57\textwidth]{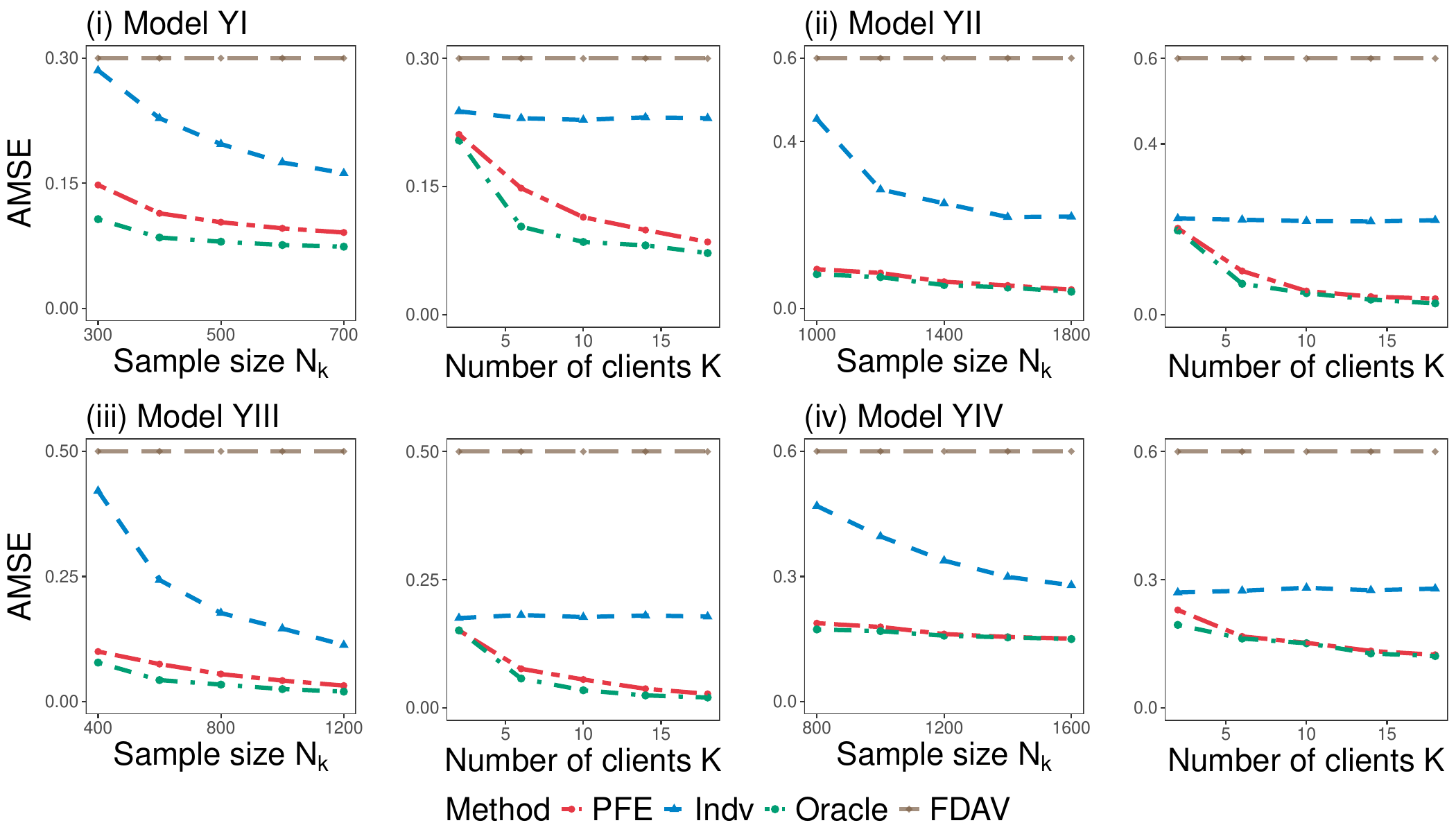}
\caption{Simulation results of AMSE varying sample size $N_k$ and the number of clients $K$ under heterogeneous scenarios.}
\label{fig:MSE_N_K}
\end{figure}

Figure~\ref{fig:MSE_N_K} further examines how AMSE changes with the local sample size $N_k$ and the number of clients $K$ under heterogeneous scenarios. For visualization, the FDAV curve is displayed at the upper boundary of each panel because its AMSE is much larger than those of the other methods. 
As $N_k$ increases, all methods improve, but PFE consistently tracks the oracle estimator and outperforms the individual estimator. As $K$ increases, the gain of PFE becomes more pronounced, since more related clients contribute to the estimation of the shared latent structure. In contrast, the individual estimator cannot benefit from additional clients, and FDAV remains affected by aggregation bias. These results confirm that PFE improves estimation efficiency through personalized federated information sharing.

\subsection{Personalized federated inference}

In this subsection, we evaluate the personalized federated inference (PFI) procedure for tail index regression coefficients. We compare PFI with individual inference (Indv) \citep{sasaki2024high}, which constructs confidence intervals using only local samples from each client. In contrast, PFI aggregates information from clients with similar tail index regression coefficients under the estimated personalized structure. We assess performance by the average confidence interval length (AL) and empirical coverage over 500 replications, with nominal level $\alpha=0.1$. 

Tables~\ref{PFI_infer_homoX} and~\ref{PFI_infer_heteroX} report the results under homogeneous and heterogeneous covariate distributions, respectively, with $N_k=2000$. In both cases, we consider the heterogeneous coefficient scenario and focus on the coefficient of the first covariate for the first client. Under homogeneous covariates, all clients follow Model XI; under heterogeneous covariates, $K/2$ clients follow Model XI and the remaining $K/2$ clients follow Model XII. The response models are Model YI--Model YIV. The threshold $w_k$ is selected according to the procedure described in Section~\ref{sec:tuning_select}.

\begin{table*}[!tb]
	\centering
 \caption{Simulation results for personalized inference under homogeneous covariate distributions.}
	\begingroup
	\setlength{\tabcolsep}{6pt} 
	\renewcommand{\arraystretch}{0.8} 
        \label{PFI_infer_homoX}
	\begin{tabular}{cccccccccccccc}
		\toprule
        &  && \multicolumn{2}{c}{$K=4$}  && \multicolumn{2}{c}{$K=8$} && \multicolumn{2}{c}{$K=12$}  \\
        \cline{4-5}  \cline{7-8} \cline{10-11} 
		 Model & Method   && AL & Coverage && AL & Coverage && AL & Coverage \\\hline
        \multirow{2}{*}{Model YI} & PFI  && 0.790 & 0.956  && 0.552 & 0.946  && 0.435 & 0.914  \\
        & Indv &&  0.968 & 0.964 && 0.968 & 0.970 && 0.968 & 0.966  \\\hline
		\multirow{2}{*}{Model YII} & PFI  &&  0.907 & 0.952   && 0.765  & 0.930 && 0.681 & 0.926  \\
        & Indv && 1.174 & 0.960  &&1.174 & 0.958 &&  1.174 &  0.962 \\\hline
		\multirow{2}{*}{Model YIII } & PFI  && 0.684 & 0.962   && 0.559  & 0.940 && 0.486 & 0.916  \\
        & Indv && 0.899 & 0.964  && 0.898 & 0.966 &&  0.898 & 0.968 \\\hline
		\multirow{2}{*}{Model YIV } & PFI  && 1.054 & 0.968   && 0.847 & 0.940   && 0.715 & 0.928  \\
        & Indv && 1.423 & 0.978  && 1.423 & 0.982  && 1.423 & 0.982   \\
		\bottomrule
	\end{tabular}
    \endgroup
\end{table*}

\begin{table*}[!tb]
	\centering
	\caption{Simulation results for personalized inference under heterogeneous covariate distributions.}
	\begingroup
	\setlength{\tabcolsep}{6pt} 
	\renewcommand{\arraystretch}{0.8} 
        \label{PFI_infer_heteroX}
	\begin{tabular}{cccccccccccccc}
		\toprule
        &  && \multicolumn{2}{c}{$K=4$}  && \multicolumn{2}{c}{$K=8$} && \multicolumn{2}{c}{$K=12$}  \\
        \cline{4-5}  \cline{7-8} \cline{10-11} 
		 Model & Method && AL & Coverage && AL & Coverage && AL & Coverage \\\hline
		\multirow{2}{*}{Model YI} & PFI && 0.597 & 0.948 && 0.449 & 0.918 && 0.371 & 0.900 \\
        & Indv && 0.755 & 0.966 && 0.757 & 0.962 && 0.757 & 0.964\\\hline
		\multirow{2}{*}{Model YII} & PFI  && 0.679 & 0.958 &&  0.518 & 0.938 && 0.471 & 0.918 \\
        & Indv && 0.860 & 0.974 && 0.862 & 0.976 && 0.860 & 0.976\\\hline
		\multirow{2}{*}{Model YIII} & PFI && 0.515 & 0.964 && 0.409 & 0.938 && 0.384 & 0.908 \\
         & Indv && 0.693 & 0.962 && 0.694 & 0.966 && 0.694 & 0.966\\\hline
		\multirow{2}{*}{Model YIV} & PFI  && 1.054 & 0.968 && 0.842 & 0.938 && 0.706 & 0.922\\
        & Indv && 1.421 & 0.978 && 1.419 & 0.980 && 1.419 & 0.980\\
		\bottomrule
	\end{tabular}
    \endgroup
\end{table*}

\begin{figure}[!tb]
\centering
\includegraphics[scale=0.72]{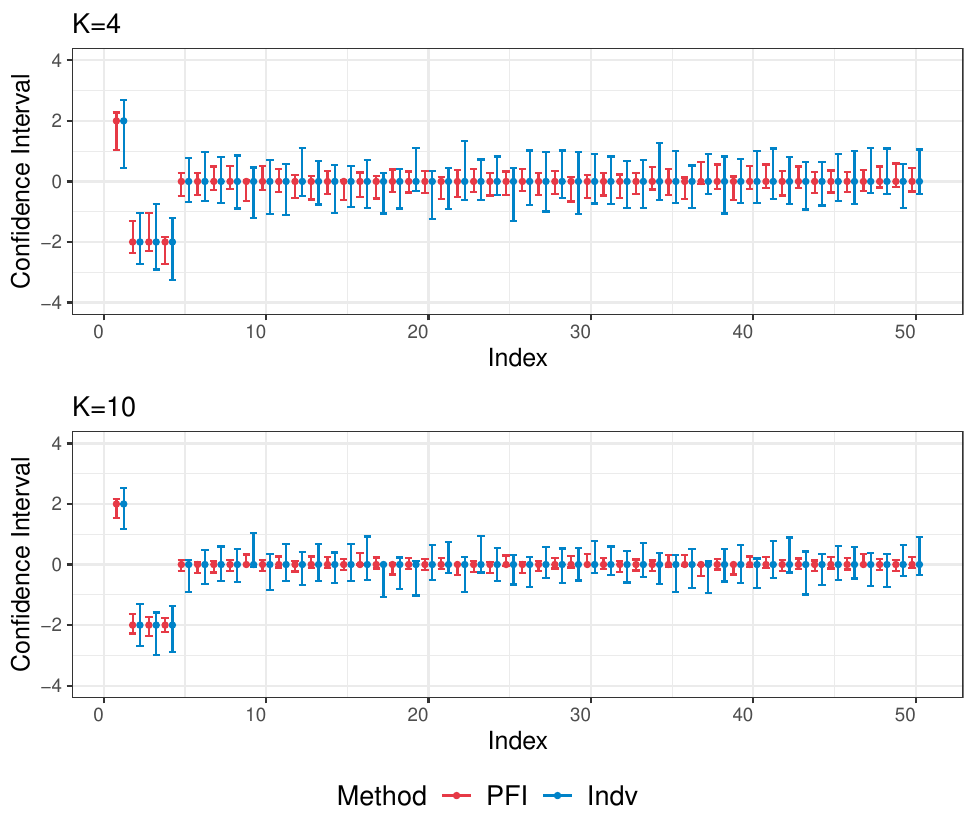}
\caption{Comparison of confidence intervals produced by PFI and Indv under Model XI YI.}
\label{fig:CI}
\end{figure}

Across all settings, PFI produces substantially shorter confidence intervals than Indv while maintaining empirical coverage at or above the nominal level. As $K$ increases, the average length of PFI intervals decreases steadily, reflecting the efficiency gain from aggregating information across related clients. By contrast, the interval length of Indv remains largely unchanged because it relies only on local samples. 
Figure~\ref{fig:CI} gives a visual comparison: PFI intervals are shorter and more concentrated around the true parameter value, whereas Indv intervals are wider. These results show that PFI improves inferential efficiency while preserving valid uncertainty quantification under both homogeneous and heterogeneous covariate distributions.

\section{Real-data analysis}\label{sec:real_data}

We apply the proposed method to firm-level financial data from the Center for China Economic Research (CCER) Database \citep{wang2009tail}. The data consist of annual reports from eight industries: Healthcare, Pharmaceuticals, Electronic Equipment, Semiconductors, Machinery Manufacturing (General Equipment), Machinery Manufacturing (Specialized Equipment), Construction and Engineering, and Real Estate. Each industry is treated as one client, so $K=8$. The response $y^{(k)}$ is the absolute return on equity (ROE), and the covariates $\bx^{(k)}$ are firm-level financial indicators. Our goal is to study how these indicators affect the tail behavior of ROE within each industry. Since firms from different industries may share common risk factors while retaining industry-specific characteristics, a personalized federated approach is natural: it borrows information across related industries while preserving cross-industry heterogeneity.

\begin{figure}[!tb]
\centering
\includegraphics[scale=0.62]{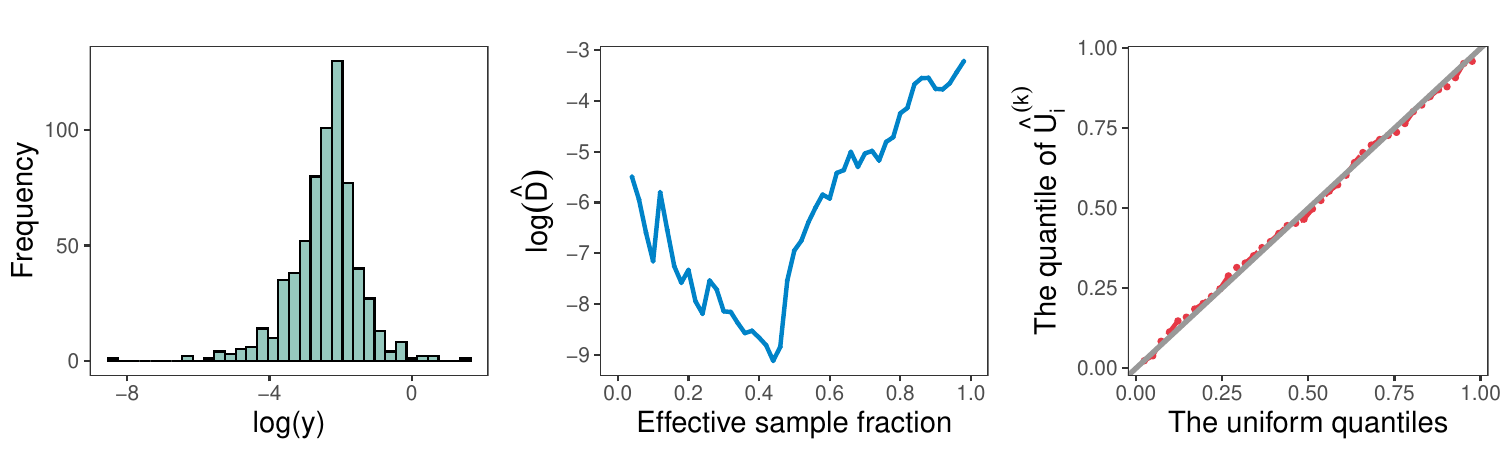}
\caption{Diagnostic plots for the real-data analysis. Left: histogram of $\log(y)$, showing heavy-tailedness of ROE. Middle: $\log(\widehat D)$ as a function of the sample fraction $n_k/N_k$. Right: QQ plot of $\widehat{U}_i^{(k)}$ against the uniform distribution for the first client under the selected $w_k$.}
\label{fig:realdata}
\end{figure}

We begin with 41 financial indicators as candidate covariates. To reduce multicollinearity among accounting variables, we remove highly collinear variables and retain 21 indicators for analysis. Since financial indicators may themselves be heavy-tailed, we apply a rank transformation before fitting the tail index regression model. Specifically, for the $j$-th covariate on client $k$, we set
$x_{ij}^{(k)}=\sqrt{12}\left(R_{ij}^{(k)}/N_k-\frac{1}{2}\right),$
where $R_{ij}^{(k)}$ is the rank of the $i$-th observation among all observations of the $j$-th covariate on client $k$. This transformation reduces the influence of extreme covariate values and places different financial indicators on a comparable scale. Thus, the estimated coefficients should be interpreted as effects of rank-based financial predictors on ROE tail behavior, rather than marginal effects on the original measurement scale. For each client, the threshold $w_k$ and hence the effective sample size $n_k$ are selected using the procedure in Section~\ref{sec:tuning_select}. Figure~\ref{fig:realdata} summarizes the diagnostic procedure, and Table~\ref{efficetive sample size} reports the original sample size $N_k$ and the effective sample size $n_k$ for each industry.

\begin{table*}[!tb]
	\centering
	\caption{Sample size $N_k$ and effective sample size $n_k$ on each client}
	\begingroup
	\setlength{\tabcolsep}{6pt} 
	\renewcommand{\arraystretch}{0.8} 
        \label{efficetive sample size}
	\begin{tabular}{cccccccccccccc}
		\toprule
		  & Heal & Phar & Elec & Semi & Mach-ge & Mach-sp & Arch & Real  \\\hline
        $N_k$ & 657 & 1754 & 2168 & 495 & 1217 & 1319 & 681 & 772 \\
        $n_k$ & 289 & 350  & 130  & 148 & 267  & 184  & 286 & 108 \\
		\bottomrule
	\end{tabular}
    \endgroup
\end{table*}

To evaluate estimation performance, we randomly split the observations in each client into training and test sets, using two-thirds for training and one-third for testing. The coefficients $\widehat{\bth}_k$ are estimated on the training set, and prediction errors are computed on the test set as
$$
\text{Prediction Error} = \frac{1}{n_{\text{test}}} \sum_{k=1}^K \sum_{i=1}^{n_{\text{test}}^{(k)}} 
\big\{\exp{(\bx_i^{(k)\T}\widehat{\bth}_k)} - \log(y_i^{(k)}/w_k)\big\}^2,
$$
where $n_{\text{test}}^{(k)}$ denotes the effective sample size in the test set for client $k$, and $n_{\text{test}} = \sum_{k=1}^K n_{\text{test}}^{(k)}$ is the total test sample size.

\begin{figure}[!tb]
\centering
\includegraphics[scale=0.6]{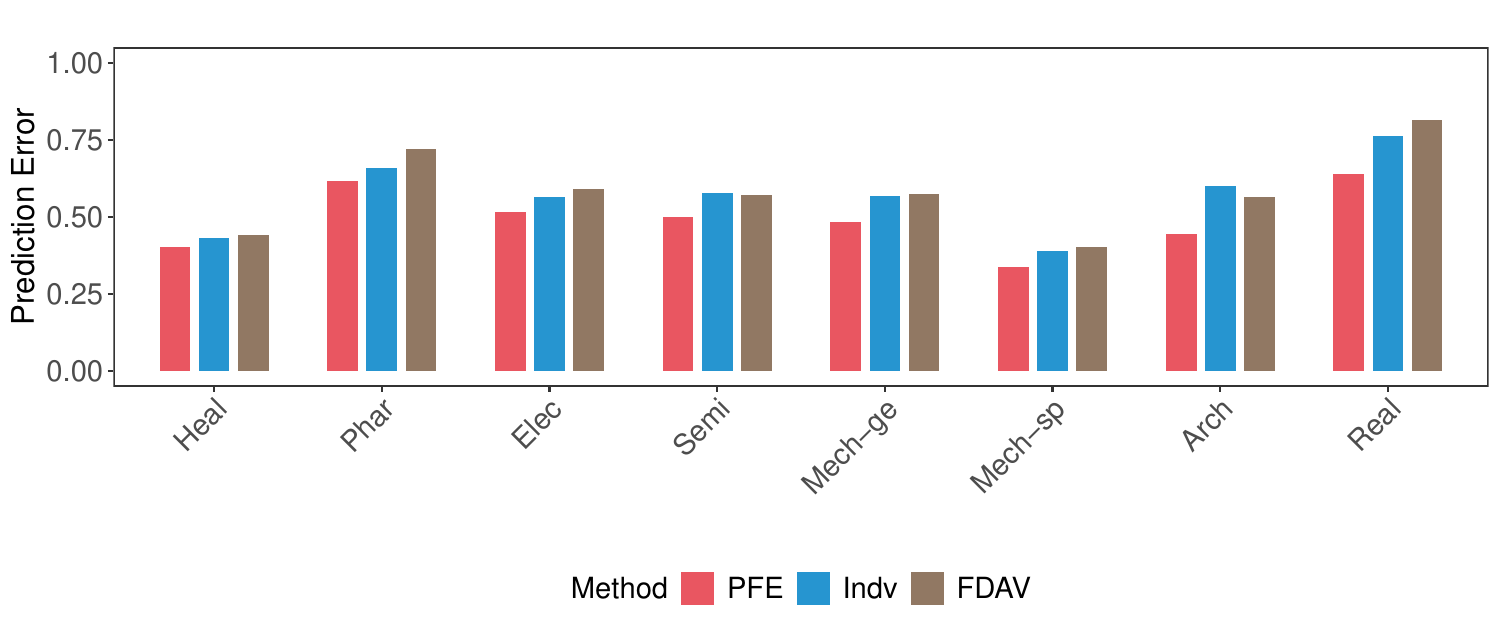}
\caption{Prediction errors of PFE, Indv, and FDAV on the test set across the eight clients.}
\label{fig:realdata OPE}
\end{figure}

\begin{table*}[!tb]
\centering
\small
\caption{Significant variables ($p<0.05$) and their CI lengths. Variables abbreviations: INT = Intercept, LEV = Leverage, FEX = Financial Expense, PUC = Paid-up Capital, STB = Short-term Borrowing.}
\begingroup
\setlength{\tabcolsep}{4pt}
\renewcommand{\arraystretch}{0.8}
\label{tab:ROE_tail_index_three_decimal}
\begin{tabular}{cccccc}
\toprule
Client & Method & Variable & $p$-value & CI length \\\hline

 \multirow{2}{*}{Heal}  & PFI & INT, LEV, FEX, PUC & 0.000, 0.000, 0.047, 0.022 & 0.228, 0.400, 0.232, 0.572 \\
  & Indv & INT & 0.000 & 0.519 \\\hline

\multirow{2}{*}{Phar} & PFI & INT, LEV, FEX & 0.000, 0.000, 0.047 & 0.228, 0.400, 0.232 \\
  & Indv & INT, LEV, STB & 0.000, 0.001, 0.039 & 0.501, 0.899, 0.569 \\\hline

\multirow{2}{*}{Elec}  & PFI & INT, LEV, FEX & 0.000, 0.000, 0.047 & 0.228, 0.400, 0.232 \\
 & Indv & INT & 0.000 & 1.485 \\\hline

\multirow{2}{*}{Semi}  & PFI & INT, LEV, FEX & 0.000, 0.000, 0.047 & 0.228, 0.400, 0.232 \\
 & Indv & INT & 0.000 & 0.779 \\\hline

\multirow{2}{*}{Mach-ge}  & PFI & INT, LEV, FEX, PUC & 0.000, 0.000, 0.047, 0.022 & 0.228, 0.400, 0.232, 0.572 \\
 & Indv & INT & 0.000 & 0.582 \\\hline

\multirow{2}{*}{Mach-sp}    & PFI & INT, LEV, FEX, PUC & 0.000, 0.000, 0.047, 0.022 & 0.228, 0.400, 0.232, 0.572 \\
 & Indv & INT & 0.000 & 1.178 \\\hline

\multirow{2}{*}{Arch}   & PFI & INT, LEV, FEX & 0.000, 0.000, 0.047 & 0.228, 0.400, 0.232 \\
 & Indv & INT, LEV & 0.000, 0.024 & 0.362, 0.727 \\\hline

\multirow{2}{*}{Real}  & PFI & INT, LEV, FEX & 0.000, 0.000, 0.047 & 0.228, 0.400, 0.232 \\
 & Indv & - & - & - \\

\bottomrule
\end{tabular}
\endgroup
\end{table*}

Using the SCAD penalty, we compare PFE with the individual estimator (Indv) and the federated averaged estimator (FDAV) described in Section~\ref{sec:simu}. Figure~\ref{fig:realdata OPE} reports the prediction errors across clients. PFE achieves the smallest prediction error in all industries, indicating that adaptive information sharing across related industries improves prediction performance. In contrast, Indv relies only on local industry-specific samples, while FDAV ignores industry heterogeneity and may suffer from aggregation bias.

\begin{figure}[!tb]
\centering
\includegraphics[scale=0.63]{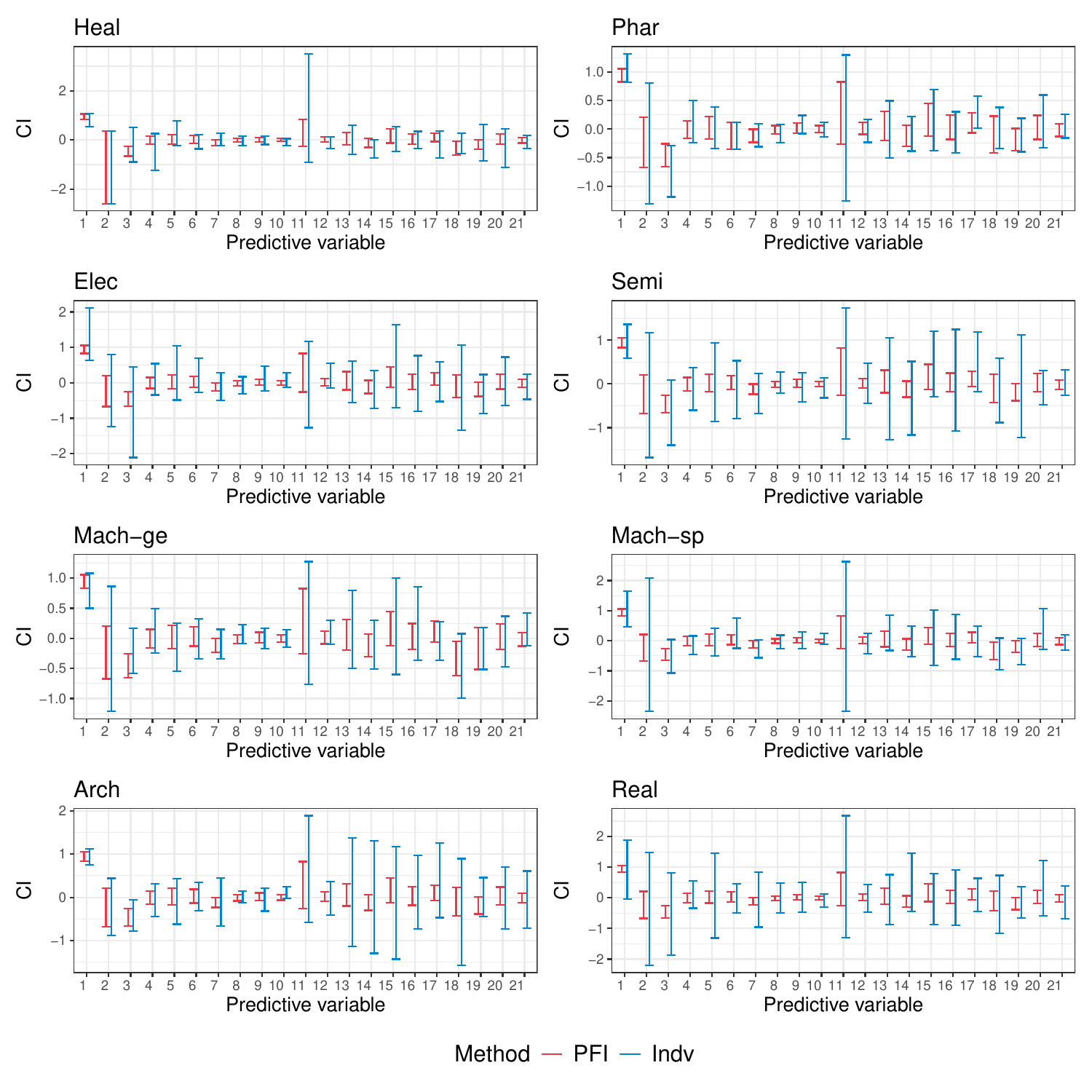}
\caption{Confidence intervals obtained by PFI and Indv on each client.}
\label{fig:realdata CI}
\end{figure}

Table~\ref{tab:ROE_tail_index_three_decimal} reports the significant variables selected by personalized federated inference (PFI) and individual inference (Indv), together with their $p$-values and confidence interval lengths. Across industries, PFI consistently identifies leverage (LEV) and financial expense (FEX) as important, suggesting that debt burden and financing cost play central roles in ROE tail behavior. Economically, high leverage increases financial risk exposure, and large financial expenses reduce net profitability, both of which may amplify extreme ROE outcomes.
In addition, PFI captures industry-specific effects. For example, paid-up capital (PUC) is selected for Healthcare and Machinery Manufacturing (both general and specialized equipment). In these sectors, higher paid-in capital may indicate greater financial flexibility and investment capacity, which can shape extreme ROE outcomes under profitable opportunities or adverse shocks. This suggests that PUC affects ROE tail behavior in a sector-specific manner, reflecting heterogeneity in capital structure and investment intensity across industries.

Compared with Indv, PFI produces more stable and interpretable results. It yields shorter confidence intervals and more consistent variable selection patterns across clients, whereas Indv often selects only the intercept or produces wider intervals due to limited tail observations. Figure~\ref{fig:realdata CI} further shows that PFI intervals are generally shorter and more concentrated around the estimated coefficients, highlighting its improved inference efficiency and its ability to identify both shared and industry-specific predictors of extreme ROE outcomes.

\section{Conclusion and discussion}

In this paper, we developed a personalized federated framework for high-dimensional tail index regression under heterogeneous client structures. The proposed method combines sparsity regularization with nonconcave fusion penalties to perform coefficient estimation, variable selection, and latent group recovery while preserving client-specific tail behavior. We established non-asymptotic convergence rates and showed that the estimator enjoys an oracle property by consistently recovering the underlying grouping structure. We further developed a debiased federated inference procedure based on adaptive weighted aggregation across related clients, leading to valid confidence intervals and hypothesis tests. Theoretical and empirical results demonstrate that the proposed method improves estimation and inference efficiency over target-only analysis.

Several extensions are worth pursuing. First, the current framework focuses on sparse linear tail index regression with a log-link specification; extending it to semiparametric, nonlinear, or varying-coefficient tail index models would broaden its applicability. Second, more complex cross-client structures, such as low-rank, network, or hierarchical relationships, could be incorporated beyond coefficient-level grouping. Finally, the framework may be extended to other extreme-value problems, including multivariate extremes and conditional risk measures.

\subsection*{Data availability statement}

The Center for China Economic Research (CCER) dataset analyzed in this study is available at \url{https://www.ccerdata.cn/}.

\subsection*{Supplementary material}

The supplementary material contains the proofs of the theorems and propositions, as well as some additional numerical results.


\vspace{0.5cm}
{\small \baselineskip 10pt
\bibliographystyle{asa}
\bibliography{TR}
}

\end{document}